\documentclass[twocolumn,tighten, twocolappendix]{aastex63}
\usepackage{longtable}
\usepackage{booktabs}
\usepackage{tabularx}
\usepackage{graphicx}
\usepackage{here}

\newcommand{\argmin}{\mathop{\rm argmin}\limits} 

\accepted{Sep 30, 2021}

\submitjournal{ApJ}

\shortauthors{Yamaguchi et al.}

\definecolor{MyDarkBlue}{rgb}{0,0.08,0.5}
\definecolor{MyDarkRed}{rgb}{0.7,0.02,0.02}
\definecolor{MyDarkmagenta}{rgb}{0.0,0.7,0.0}
\newcommand{\utokyo}{Department of Astronomy, Graduate School of Science, The University of Tokyo, 7-3-1 Hongo, Bunkyo-ku, Tokyo 113-0033, Japan}
\newcommand{\naoj}{National Astronomical Observatory of Japan, 2-21-1 Osawa, Mitaka, Tokyo 181-8588, Japan}
\newcommand{\ism}{The Institute of Statistical Mathematics, 10-3 Midori-cho, Tachikawa, Tokyo 190-8562, Japan}
\newcommand{\sokendaiism}{Department of Statistical Science, School of Multidisciplinary Sciences, Graduate University for Advanced Studies, 10-3 Midori-cho, Tachikawa, Tokyo 190-8562, Japan}
\newcommand{\sokendainaoj}{Department of Astronomical Science, School of Physical Sciences, Graduate University for Advanced Studies, 2-21-1 Osawa, Mitaka, Tokyo 181-8588, Japan}
\newcommand{\kogakuin}{Division of Liberal Arts, Kogakuin University, 1-24-2 Nishi-Shinjuku, Shinjuku-ku, Tokyo 163-8677, Japan}
\newcommand{\abc}{Astrobiology Center, National Institutes of Natural Sciences, 2-21-1 Osawa, Mitaka, Tokyo 181-8588, Japan}
\RequirePackage[normalem]{ulem}
\RequirePackage{color}\definecolor{RED}{rgb}{1,0,0}\definecolor{BLUE}{rgb}{0,0,1} 
\begin{document}
%-------------------------------------------
-----------------------------------
% Title
%------------------------------------------------------------------------------
\title{ALMA Super-resolution Imaging of T Tau:\\ r = 12 au Gap in the Compact Dust Disk around T Tau N}

%------------------------------------------------------------------------------
% Authors
%------------------------------------------------------------------------------
\correspondingauthor{Masayuki Yamaguchi}
\email{masayuki.yamaguchi.astro@gmail.com}
\author{Masayuki Yamaguchi}
\affil{\utokyo}
\affil{\naoj}

\author{Takashi Tsukagoshi}
\affil{\naoj}
\author{Takayuki Muto}
\affil{\kogakuin}
\author{Hideko Nomura}
\affil{\naoj}
\affil{\sokendainaoj}

\author{Takeshi Nakazato}
\affil{\naoj}
\author{Shiro Ikeda}
\affil{\ism}
\affil{\sokendaiism}
\author{Motohide Tamura}
\affil{\utokyo}
\affil{\naoj}
\affil{\abc}
\author{Ryohei Kawabe}
\affil{\naoj}
\affil{\sokendainaoj}
%
%\author{Others}
%
%------------------------------------------------------------------------------
% Abstract
%------------------------------------------------------------------------------ 
\begin{abstract}
 
Based on Atacama Large Millimeter/submillimeter Array (ALMA) observations, compact protoplanetary disks with dust radii of $r\lesssim 20-40$ au were found to be dominant in nearby low-mass star formation regions. However, their substructures have not been investigated because of the limited spatial resolution achieved so far. We apply a newly developed super-resolution imaging technique utilizing sparse modeling (SpM) to explore several au-scale structures in such compact disks. SpM imaging can directly solve for the incomplete sampling of visibilities in the spatial frequency and potentially improve the fidelity and effective spatial resolution of ALMA images. Here, we present the results of the application to the T Tau system. We use the ALMA 1.3 mm continuum data and achieve an effective spatial resolution of $\sim 30\%$ (5 au) compared with the conventional CLEAN beam size at a resolution of 17 au. The reconstructed image reveals a new annular gap structure at $r= 12$ au in the T Tau N compact disk with a dust radius of 24 au, and resolves the T Tau Sa and Sb binary into two sources. If the observed gap structure in the T Tau N disk is caused by an embedded planet, we estimate a Saturn-mass planet when the viscous parameter of the disk is $10^{-3}$. Ultimately, ALMA observations with enough angular resolution and sensitivity should be able to verify the consistency of the super-resolution imaging and definitely confirm the existence of this disk substructure.
 
\end{abstract}

%%%%%%%%%%%%%%%%%%%%%%%%%%%%
%%%% Start of document %%%%% 
%%%%%%%%%%%%%%%%%%%%%%%%%%%%

%------------------------------------------------------------------------------
%Introduction
%------------------------------------------------------------------------------
\section{Introduction}\label{sec:1}
Planets are formed in protoplanetary disks (PPDs) around young stars, which are composed of gas and dust \citep[e.g., ][]{Hayashi1985}. The structure and evolution of PPDs are thought to be closely linked to the formation process of planets for both core accretion and disk instability models \citep[e.g.,][]{Johansen2007, Ida2013}. Protoplanets with a sufficiently large mass can induce the formation of a gap in the the disks \citep[e.g.,][]{Lin1986, Takeuchi1996, Zhu2012, Pinilla2012b}. The minimum gap-opening mass depends on the viscosity and scale-height of the disk, and ideally super-Earth mass planets ($\sim 10~M_{\bigoplus}$) can produce detectable gaps in (sub-)millimeter regime \citep[][]{Rosotti2016, zhang2018}. Once the gap is spatially resolved, its width and depth can be used to estimate the mass of a growing planet \citep[e.g., ][]{Kanagawa2015, Kanagawa2016, zhang2018}. Meanwhile, several alternative explanations have been proposed for the origin of the gap; e.g., the effect of the snow lines of major volatiles \citep[e.g.,][]{zhang2015, Okuzumi2016}, magneto-hydrodynamic effects \citep[e.g.,][]{Flock2015}, secular gravitational instability \citep[e.g.,][]{Youdin2011,Takahashi2014, Takahashi2016}, and thermal wave instability \citep[e.g.,][]{Watanabe2008, Siebenmorgen2012, Ueda2019,Ueda2021}. Effectiveness of each proposed mechanism depends on the physical and chemical properties of the disks. The different mechanisms may work together in some disks, or a dominant mechanism may differ among disks. Understanding the origin of substructures, such as gaps and rings, and their links to planet formation is currently one of the hot topics in the field.

The advent of the Atacama Large Millimeter/submillimeter Array (ALMA) has enabled us to observe PPDs with high spatial resolution, and transformational images or analysis (e.g., interferometric modeling) of disks have been produced. For instance, the disk substructures at the High Angular Resolution Project (DSHARP) and the Ophiuchus DIsc Survey Employing ALMA (ODISEA) provided ALMA images for bright and large disks with radii of $r = 50-260$ au with angular resolutions down to $2 - 5$ au \citep[for consistency, we always refer to ``dust disk'' as ``disk'';][]{Andrews2018DSHARP, cieza2020}. These results reveal an annular gap structure, which are likely carved by a planet with Neptune-Jupiter mass \citep[][]{zhang2018}. However, disks of small sizes ($r\lesssim 20 - 40$ au) of PPDs were found to be dominant in the fraction ($\sim70 - 90\%$) in low-mass star-forming regions \citep{Cieza2019, Long2019, Ansdell2016}, but their substructures have not been well investigated. Small disks are typically less massive in terms of the disk mass compared with large disks, and will be key to investigating the missing link between PPD substructures such as gaps and their locations. There is the need for extensive research on such a major PPD population to investigate the inner $r = 5 - 40$ au region in such PPDs as a possible location for the formation of giant planets \citep[e.g.,][]{Bate2018,Lodato2019}

To explore a few au-scale gap structures in such compact disks in nearby low-mass star formation regions, a high spatial resolution of $\lesssim 0''.035$ is required to resolve, e.g., a gap formed by a Jupiter-mass planet orbiting around a low-mass star ($0.5~ M_{\odot}$, $d= 140$ pc) where the gap width is assumed to be roughly 5 au and calculated to be $5.5$ times of Hill radius at $ r = 10$ au \citep[][]{Lodato2019}. The highest angular resolution in ALMA Band 6 observations achieved thus far is  $\sim 0''.02 - 0''.05$ (e.g., DSHARP and ODISEA). ALMA high-resolution observations potentially resolve the compact disk's substructure as well with sufficient $uv-$coverage by longer observing time  \citep[e.g., SR 4, DoAr 33, and WSB 52;][]{huang2018}. Sparse modeling (SpM) is another approach, i.e., a promising technique that can achieve such a high spatial resolution, even in the lower frequency ALMA Band 4 and 6. This technique has already been applied to the imaging of the event horizon telescope (EHT) \citep{eht2019d} and ALMA \citep{yamaguchi2020, Aizawa2020}. To date, the use of EHT mock observational data and ALMA actual observational data has confirmed that this technique achieves a higher-fidelity image than the conventional CLEAN algorithm at the angular scale of $30\% - 40\%$ of the CLEAN beam \citep[i.e., super-resolution;][]{honma2014, akiyama2017a, akiyama2017b,Kuramochi2018, yamaguchi2020}. Furthermore, with an emphasis on improving the fidelity even in super-resolution regimes and at the calculation speeds, a new SpM imaging software intended for ALMA observational data has been developed over the last several years \citep{nakazato2020}.

Here, we focus on the PPD around the T Tau triple star system. T Tau is a triple star that became an eponymous member of the class of low-mass, pre-main sequence stars \citep{Joy1945}. This system consists of a star (T Tau N) in the north and a close binary (T Tau Sa and Sb) in the south \citep{Dyck1982, Koresko2000}, located in the Taurus star-forming region at a distance of $143.7\pm 1.2$ pc, as measured by Gaia DR2 \citep{Gaia2016, Gaia2018}. Both T Tau N and T Tau S (Sa $+$ Sb) are embedded in an infalling envelop \citep[][]{Momose1996}, jets have been found to associated with both sources \citep[][]{Beck2020}. T Tau N is classified as Class II, while T Tau S is a Class I system \citep[][]{Furlan2006, Luhman2010}. The mass of T Tau Sa and Sb are $2.1~M_{\odot}$ and $0.4~M_{\odot}$, respectively \citep{Schaefer2020}. T Tau N is one of the brightest classical T Tauri stars in Taurus. The stellar properties of T Tau N have been calculated using optical spectral types combined with stellar evolutionary models in several studies, and we adopted a stellar bolometric luminosity of $6.82~L_{\odot}$ and a stellar mass of $2.19~M_{\odot}$ \citep{Herczeg2014}. 

Previous ALMA continuum observations targeted the dust disk around the T Tau system, but were unable to spatially resolve them on the CLEAN image because of the insufficient spatial resolution of $0''.12$ (or 17 au) \citep{Long2019}.  The T Tau Sa/Sb disk is only seen as a single Gaussian-like distribution. The T Tau N disk was found to be a bright disk with a total flux of $\sim180$ mJy at 1.3 mm, but is only seen as a flat compact disk with a radius of $\sim 20$ au. Similarly, neither ground-based near-infrared adaptive optics observations nor space-based optical observations can resolve the disk around T Tau N and T Tau Sa/Sb well with a resolution of $\sim0''.07$ or 10 au \citep[e.g.,][]{Yang2018}. Intriguingly, \cite{Manara2019} pointed out that significance residuals ($\sim3\sigma$) were found both at the T Tau N and T Tau S after subtracting axisymmetric models of the two sources from the ALMA continuum image. It can be interpreted as tentative evidence of disk substructures around the T Tau system. Such bright and compact disks around T Tau system would be most suitable for exploring several au-scale structures using SpM imaging.

In this study, by using super-resolution imaging with SpM, which is an approach that has been proven in previous studies, we present a high-resolution (5 au or $0''.03$) image of the T Tau system. We find an annular gap at $r=12$~au in the disk around T Tau N and two separate point-like dust continuum emissions, which are located at positions corresponding to T Tau Sa and Sb. In Section \ref{sec:data_reduction}, we describe the data reduction and imaging with both CLEAN and SpM. In Section \ref{sec:results}, we show the resulting images of a 1.3 mm continuum emission, and present the findings of the substructure of the T Tau N disk and the two disks originating from T Tau Sa and Sb on the SpM image. In section \ref{sec:discussion}, we discuss the expected origins of the annular gap found in the T Tau N disk.

\section{Data Reduction and Imaging}\label{sec:data_reduction}
\subsection{Data Reduction and Imaging with CLEAN}\label{sec:2.1}

We reanalyzed the ALMA archival data obtained for T Tau on August 18, 2017, as part of the project 2016.1.01164.S (PI: Herczeg), including the continuum at 225.5 GHz and $\rm ^{13}CO~(J = 2-1)$ and $\rm C^{18}O~(J = 2-1)$ line data. Continuum data have already been published in \cite{Long2019, Manara2019, Beck2020}. The observations were performed with a 12-m array consisting of forty-three 12-m antennas (C40-7 antenna configuration with the baseline length extending from 21.0 m to 3637.7 m) and the on-source time of the target source was 8 min.

The data consisted of four spectral windows (spws). Two of the spws were used for the continuum observations and had center frequencies of 218 and 233 GHz. The average observation frequency was 225.5 GHz (wavelength of 1.3 mm). The other spws were used to cover $\rm ^{13}CO$ and $\rm C^{18}O$ with a velocity resolution of $\rm 0.16~km~s^{-1}$. In this study, we used continuum spws to reconstruct images by employing two different techniques, namely CLEAN and SpM. The $\rm ^{13}CO$ and $\rm C^{18}O$ data were analyzed, but emissions associated with T Tau S and N were not identified in the two lines.

The raw data were calibrated using the Common Astronomy Software Applications package \citep[\tt CASA;][]{mcmullin2007}, version 5.1.1. The initial calibration was performed using the ALMA pipeline on $\tt CASA$. In the pipeline, $\rm J0423 - 0120$ was used for the flux and bandpass calibration, and $\rm J0431+1731$ was used for phase calibration. The positional offset between the phase (map) center and the emission peak of the T Tau N disk was adjusted using the $\tt CASA$ task $\tt fixvis$.

The data were firstly imaged with the $\tt tclean$ task (hereafter CLEAN) by adopting Briggs weighting ($\tt robust = 0.5$). The CLEAN is the most standard image reconstruction algorithm and also one of the nonlinear deconvolution technique \citep[e.g., ][]{hogbom1974, clark1980, schwab1984, cornwell2008, Zhang2020clean}. The technique iteratively determines the point source on the image domain that best fits the observed visibilities, starting from a dirty image, which is obtained by the Fourier transform of the observed visibility with non-observed data filled with zero. This process is repeated until some convergence requirement is met. The final image is obtained by convolving the point-source model (CLEAN model) with an idealized CLEAN beam (usually an elliptical Gaussian fitted to a synthesized beam). We note that the beam-convolution in the image domain corresponds to multiplication in the visibility domain, which causes a loss in spatial resolution in the visibility domain via an underestimate of the observed visibility amplitudes (see Appendix \ref{appendix:images_comparison}).

Next, to improve the signal-to-noise ratio (SNR) of the image by correcting a systematic gain error (e.g., antenna-based and baseline-based errors), we performed two rounds of phase (longer at the 1st (98 s) and down to the integration time at the 2nd (49 s) with $\tt calmode = p$) and one round of amplitude and phase (integration time of 98 s with $\tt calmode = ap$) self-calibrations. We obtained the final CLEAN image (= CLEAN model convolved with CLEAN beam $+$ residual map) after self-calibration with the signal-to-noise ratio improved by a factor of 3.8, compared with the initial one. The CLEAN beam size $\theta_{\rm CLEAN}$ was $0''.14 \times 0''.10$ at PA of $34^{\circ}.1$, and its peak intensity and RMS noise level (collected noise values for $r > 3''.0$ from the phase center) were $63.81~\rm mJy~beam^{-1}$ and $41~ \rm \mu Jy~beam^{-1}$, respectively. These values are in relatively good agreement with those reported previously in \cite{Long2019} (i.e., peak intensity = $64.56~\rm mJy~beam^{-1}$, RMS noise = $52~ \rm \mu Jy~beam^{-1}$).

\subsection{Imaging with Sparse Modeling}\label{sec:ImagingwithSpM}
We performed the SpM imaging  \citep{yamaguchi2020}. Here, we briefly describe the outline of the SpM imaging and the cross validation (CV), which were used for the imaging.

The self-calibrated visibility data were adopted for the image reconstruction with the latest SpM imaging task, $\tt PRIISM$, ver.0.3.0 \footnote[1]{$\tt PRIISM$ (Python Module for Radio Interferometry Imaging with Sparse Modeling) is an imaging tool for ALMA based on the sparse modeling technique, and is publicly available at \url{https://github.com/tnakazato/priism}} \citep[][]{nakazato2020} working with $\tt CASA$. The image is reconstructed by minimizing a cost function in which two convex regularization terms of the brightness distribution, $\ell_1$-norm and total squared variation (TSV), were utilized with the chi-squared error term \citep{Kuramochi2018}. The two regularizers adjust the sparsity and smoothness of the reconstructed image.  We minimize the cost function to obtain the optimum image, which is formulated as

\begin{eqnarray}\label{spm_eq}
{\bf I} & = & \argmin_{\bf I} \Big(||{\bf W}({\bf V} - {\bf FI})||^2_2 + \Lambda_l \sum_i \sum_j |\rm I_{i,j}| \nonumber \\
& + & \Lambda_{tsv} \sum_i \sum_j \big(|\rm I_{i+1,j} - \rm I_{i,j}|^2 + |\rm I_{i,j+1} - \rm I_{i,j}|^2\big)\Big),
\end{eqnarray}
\noindent
where ${{\bf I} = \{\rm I_{i,j}\}}$ is a two-dimensional (2D) image to be reconstructed, where $i$ and $j$ represent the pixel indices, ${\bf V}$ is the observed visibility (i.e., the self-calibrated visibilities), ${\bf F}$ is the Fourier matrix, and ${\bf W}$ is a diagonal weight matrix (each diagonal element is $1/ \sigma^{2}_{s}$, where $\sigma_{s}$ is the observational error of each visibility point indexed by $s$, normalizing the residual visibility $({\bf V} - {\bf FI})$ on the chi-squared term. The two regularization terms are controlled by the positive variables $\Lambda_{l}$ and $\Lambda_{tsv}$, respectively.

CV is a statistical method that is employed to choose the optimal values of regularization parameters \citep[see][for details]{akiyama2017a,akiyama2017b}. In this study, we used the 10-fold CV implemented in $\tt PRIISM$ and searched for the optimal parameter set with $5\times5$ sets of regularization parameters, which are $\Lambda_{l} = (10^{3},10^{4}, ...,10^{7}$) and $\Lambda_{tsv} =(10^{7}, 10^{8}, ..., 10^{11}$). In the imaging using $\tt PRIISM$, we used non-uniform fast Fourier transform (NuFFT) algorithms to compute the Fourier transform and perform iterative fitting of the model to visibility data ($N_{\rm iter} = 1000$) until the iteration algorithm converges.

In the process of the $N$-fold CV, the data set ${\bf V}$ is randomly divided into $N$ subsets, and $N-1$ sets are used for image reconstruction by employing the SpM imaging method with a fixed ($\Lambda_{l}$, $\Lambda_{tsv}$). The reconstructed image is then Fourier transformed, and the weighted chi-squared error, which is defined below, is computed for the remaining subset.

\begin{eqnarray}
\label{mse_eq}
 {\rm MSE} =
 \|{\bf W}  
  \left(
    {\bf V}- 
    {\bf F I}
  \right) 
 \|^2/~ {tr {\bf W}},
\end{eqnarray}
\noindent
where $tr {\bf W}$ is the trace of matrix ${\bf W}$. This process is iterated $N$ times by taking different subsets, deriving the cross-validation error (CVE) formulated in the MSE ($\Sigma^{N}_{i = 1}MSE_{i}/N$) as well as the standard deviation ($\Sigma^{N}_{i = 1}\sqrt{(MSE_{i}-CVE)^{2}}/N\sqrt{N-1}$).

We obtained 25 images corresponding to 25 different sets of ($\Lambda_{l}$ and $\Lambda_{tsv}$). The wide range of parameter space is selected via pre-tuning so that we do not miss the optimal image and so that it is possible to find it near the center of the image matrix. In this pre-tuning, $\Lambda_{l}$ is first fixed, and an optimal $\Lambda_{tsv}$ with the minimum CVE is searched in a wide range via SpM imaging. Next, the obtained optimal $\Lambda_{tsv}$ is fixed, and an optimal $\Lambda_{l}$ is similarly searched in a wide range for $\Lambda_{l}$. The values and ranges can be tuned according to the target source properties in $\tt PRIISM$. Figure \ref{fig:cvimage} shows the reconstructed images together with the calculated values of CVE. An image with the minimum CVE can be regarded as the optimal image \citep[see Appendix.\ref{appendix:cvimage};][]{akiyama2017a,akiyama2017b,Kuramochi2018, yamaguchi2020}. In other words, the image of ($\Lambda_{l}$, $\Lambda_{tsv}$) $=$ ($10^{5},10^{9}$) is selected as the optimal one.

In order to quantify the effective resolution $\theta_{\rm eff}$ of the technique, we have performed the following evaluation. We injected an artificial point source to the observed data in the visibility domain. We then performed the SpM imaging with the same regularization parameters of the optimal image as well as other sets of parameters. In the SpM images, effective resolution was evaluated with an elliptical Gaussian fit to the point source (see Figure \ref{fig:spm_resolution}). In these simulation, we refer to an evaluation of an effective spatial resolution from the non-parametric image modeling with the maximum entropy method \citep[MEM;][]{Carcamo2018, Perez2020}. The input flux density of the point source is 7.1 mJy, which is comparable with that of the emission around T Tau Sa (see Table \ref{table:ttau_propaties}). The reconstructed image for the optimal parameter provides the FWHM size (i.e., effective spatial resolution) of the point source, $\theta_{\rm eff} = 0''.038 \times 0''.027$ (or $5\times 4$ au) with a PA of $45.3^{\circ}$ and recovers a total flux of $7.9$ mJy ($\sim10\%$ higher than the input value). The obtained effective resolution is roughly consistent with the empirical values of $\sim 30\%$ of CLEAN beams $\theta_{\rm CLEAN}$ \citep[][]{akiyama2017a, akiyama2017b, Kuramochi2018,yamaguchi2020}, which is $\theta_{\rm CLEAN} = 0''.14 \times 0.''10$ with a PA of $34.1^{\circ}$. The derived $\theta_{\rm eff}$ is comparable to those in high-resolution observations ($\theta_{\rm CLEAN}\sim 0.''02-0''.05$) such as DSHARP and ODISEA even though the maximum baseline length of our data ($\sim4$ km) is $3 - 4$ times shorter than that of those high-resolution data \citep[$\sim13-16$ km;][]{Andrews2018DSHARP,cieza2020}.

The effective resolution depends on the regularization parameters, especially on the TSV term. The simulation results for other sets of regularization parameters (mainly for different $\Lambda_{tsv}$) are given in Appendix \ref{appendix: effect_resolution}; the smaller $\Lambda_{tsv}$ yet, the better spatial resolution. In Appendix \ref{appendix: effect_resolution}, we also add an interferometric theory-based probable explanation on why the SpM imaging can achieve roughly three times better spatial resolution.

%-----------------------------Start Figure------------------------------
\begin{figure*}[t]
\centering
\includegraphics[width=0.95 \textwidth]{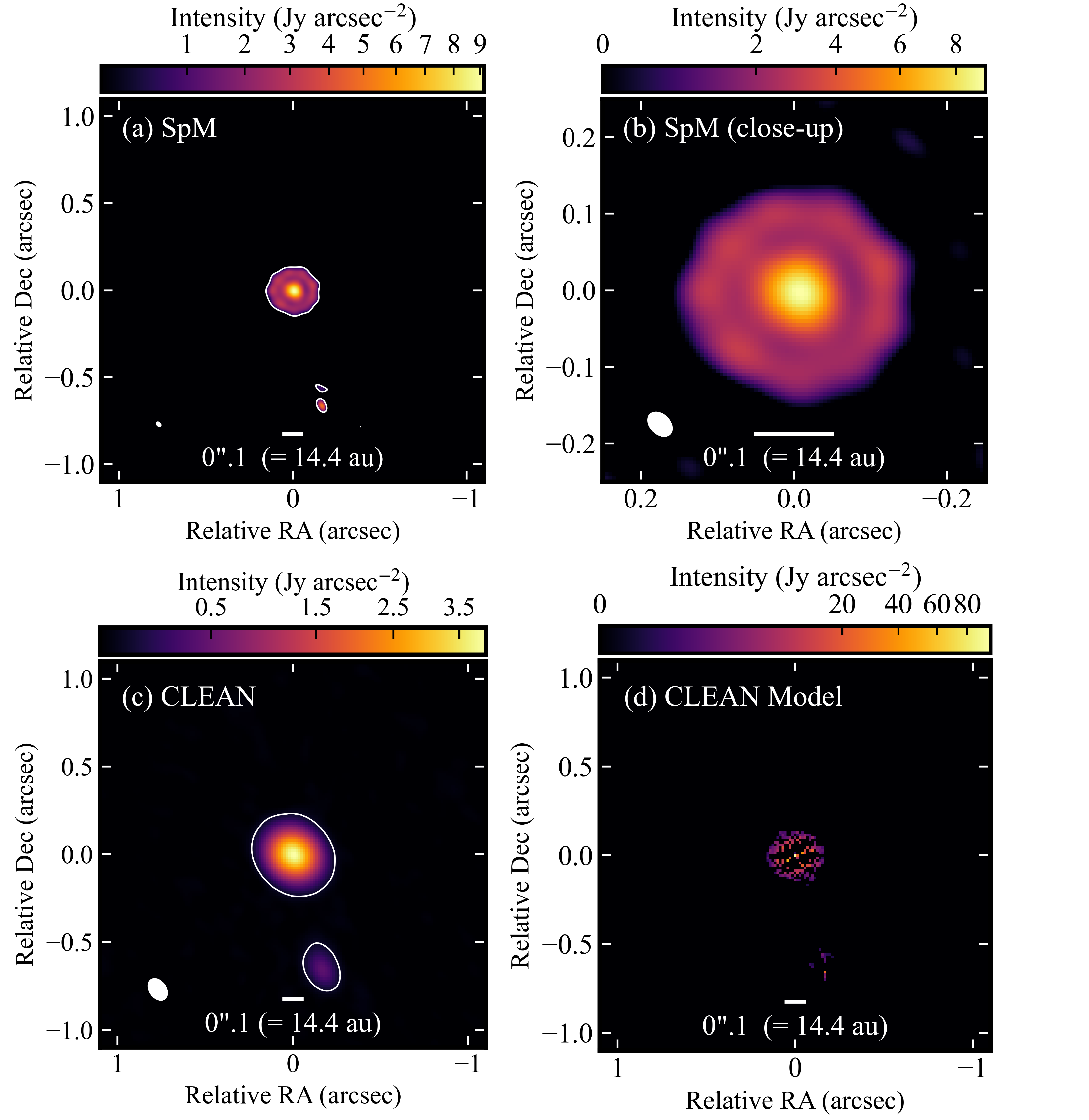}
\caption{Gallery of ALMA continuum images at 1.3 mm (Band 6) of PPD T Tau system. The same color scale given by a power law with a scaling exponent of $\gamma =  0.6$ was adopted, except for the CLEAN model image ($\gamma = 0.3$). A white bar of $0.''1$ (=14.4 au) is provided for reference to the angular scales. (a) SpM image. The filled white ellipse denotes the effective spatial resolution with a size of $0''.038 \times 0''.027$ for a PA of $45^{\circ}.3$ in the bottom left corner. The resolution is estimated from an artificial point source simulation. The contour corresponds to $I_{\rm DT}$, where $I_{\rm DT}$ is the detection threshold of $272~\rm mJy~asec^{-2}$. Note that the SpM image is not processed by a synthesized beam-convolution as a CLEAN image is done, and the unit of the SpM image is not $\rm Jy~beam^{-1}$. The unit of the SpM image that was initially obtained from the imaging is $\rm Jy~pixel^{-1}$, and we convert it to $\rm Jy~arcsec^{-2}$. (b) Close-up view centered on T Tau N of SpM image. A field of view of $0.''5\times0.''5$ is adopted. (c): CLEAN image with Briggs weighting with a robust parameter of 0.5. The filled white ellipse denotes the synthesized beam with a size of $0''.14 \times 0''.10$ for a PA of $34^{\circ}.1$ in the bottom left corner. The contour corresponds to $20\sigma_{I}$, where $\sigma_{I}$ is the RMS noise of $2.58~\rm mJy~asec^{-2}$ ($=41~\mu \rm Jy~beam^{-1}$). (d) CLEAN model image before convolution with the CLEAN beam.}
\label{fig:cont_images}
\end{figure*}
%-----------------------------End Figure---------------------------------

%-----------------------------Start Figure------------------------------
\begin{figure}[h]
\centering
\includegraphics[width=0.48 \textwidth]{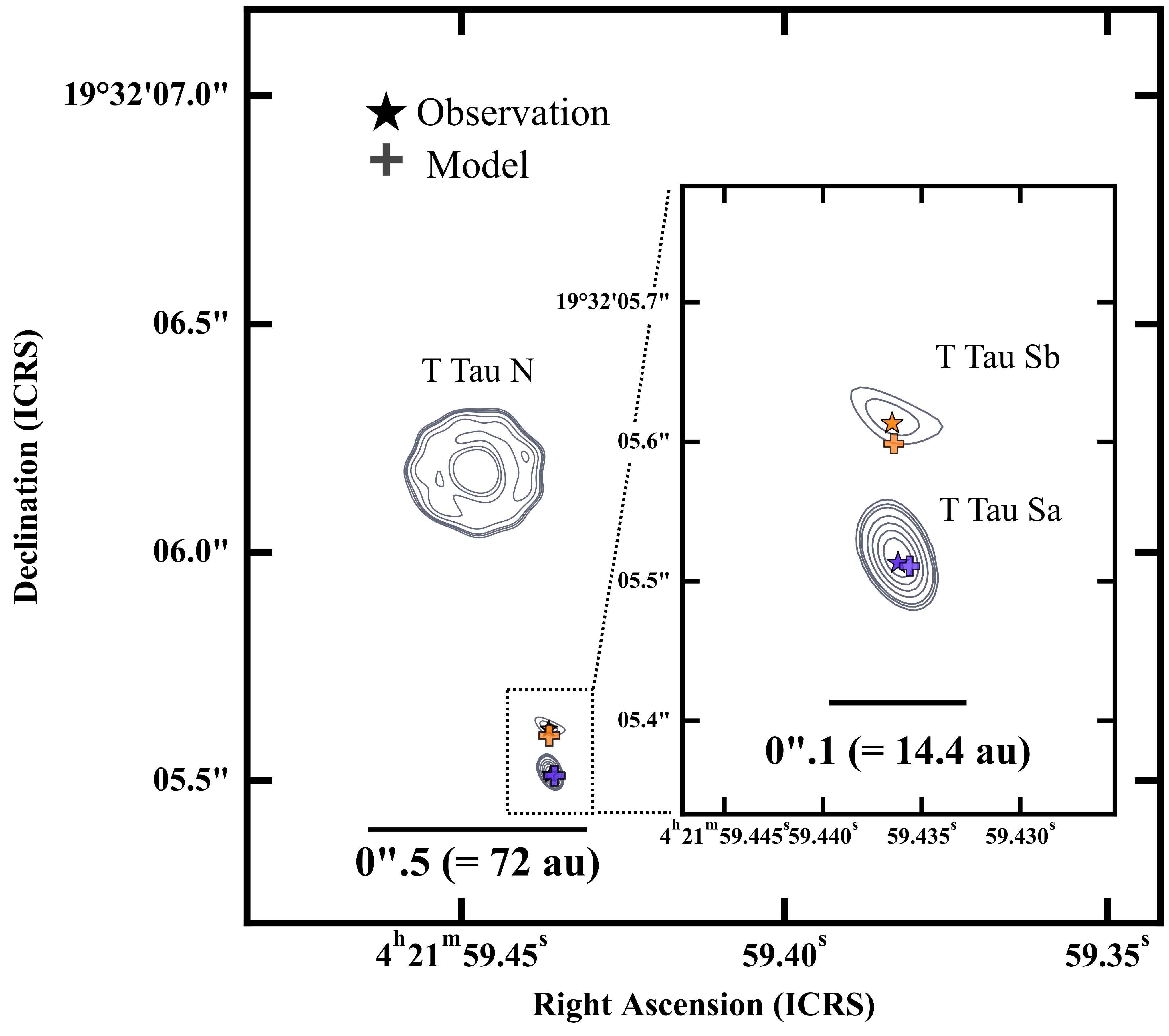}
\caption{Comparison of positions of T Tau Sa and Sb between the peak emission on the SpM image and the stellar orbit model on the date of the ALMA observations \citep[2017 August 18 UTC,][]{Kohler2016}. The positions of the sub-mm emission peaks are marked with stars, and the predicted positions of T Tau Sa and Sb are marked with crosses. These marked positions were superimposed on the SpM image. The image contours correspond to $(1,2,3,6,9,12,15) \times I_{\rm DT}$. Black bars are provided for reference to angular scales.}
\label{fig:position_TTau}
\end{figure}
%-----------------------------End Figure---------------------------------

\section{Results}\label{sec:results}
\subsection{SpM image and evaluation of its noise levels}\label{sec:spmimage}
We evaluated the noise and significance levels of the optimal SpM image. As described in \cite{yamaguchi2020}, owing to both thermal and systematic noise, an SpM image suffers from (unexpected) positive emissions in its off-source area (i.e., outside the target source area). This is because non-negative constraints have been adopted in the SpM imaging algorithm, and artificial emissions with positive intensity may be present in the off-source area. Here, we define a detection threshold (DT) in the target source area as the maximum intensity ($I_{\rm DT}$) of such artificial emissions (note that $I_{\rm DT}$ is the same definition as $I_{100}$ in \cite{yamaguchi2020}). $I_{\rm DT}$ was found to be $\rm 272~mJy~asec^{-2}$ by analyzing noise statistics at the pixel scale ($0''.05$) outside the source ($r > 0''.8$).

For direct comparison with the noise level in the CLEAN image, we convolved the optimal SpM image with the same beam size as that used for the RMS noise estimate of the CLEAN image ($\tt robust=0.5$). We found that the beam-convolved $I_{\rm DT}$ is $152~\mu\rm{Jy~beam}^{-1}$ for the optimal SpM image and was approximately 3.7 times higher than the RMS noise of the CLEAN image (see Figure \ref{fig:spm_clean}).

Another way of estimating the detection threshold is the usage of image simulation of an injected artificial point source, as described in Appendix \ref{appendix: effect_resolution}. We changed the flux density of the input point source from $1000~\mu\rm Jy$ to $100~\mu\rm Jy$ in increments of $100~\mu\rm Jy$ in the SpM simulation and judged the detection of the point source in the image. In the optimal image case (($\Lambda_{l}$, $\Lambda_{tsv}$) $=$ ($10^{5},10^{9}$)), the detection threshold is $300~\mu\rm Jy~beam^{-1}$ (see Figure \ref{fig:spm_clean}), which is two times higher than $I_{\rm DT}$. This value would provide not accurate but some reference to the threshold if we consider that the increment of the flux density is rough and the selection of the source position is not so optimized for evaluating the detection threshold precisely.

It would also be possible to estimate the noise levels with measuring the RMS noise of a residual map, which can be obtained by performing the 2D Fourier transform of residual visibilities between the SpM model data (which can be obtained by the inverse 2D Fourier transform of the SpM image) and the observed data. The residual map was reconstructed from the residual visibilities using $\tt DIFMAP$ \citep{Shepherd1994}. The residual map was created by adopting a natural $uv$ weighting, providing synthesized beams of $0''.17\times0''.12$ with a PA = $36.8^{\circ}$ and an RMS noise of $31~\mu\rm{Jy~beam}^{-1}$. This is smaller than that in the CLEAN image, and this could be because the SpM model image retrieves positive noises. For comparison, we convolved the optimal SpM image with the same beam size as that used for the RMS noise estimate. We found that the beam-convolved $I_{\rm DT} (= 152~\mu\rm{Jy~beam}^{-1})$ was approximately five times higher than the RMS. Based on the above evaluations, emission features above $I_{\rm DT}$ are considered significant in the SpM image.

Figure \ref{fig:cont_images} shows the SpM and CLEAN images of the T Tau system. The SpM image spatially resolves the disk structure around T Tau N, and an annular gap structure has been newly found. The emission around the T Tau S system is spatially resolved into two sources, although the CLEAN image does not resolve them. Table \ref{table:ttau_propaties} shows that the total fluxes of these sources obtained from the SpM image above the $I_{\rm DT}$ level are generally consistent with the values obtained from the CLEAN image above $5\sigma$ levels. SpM reproduces a high fidelity image that better fits the observed visibilities than the CLEAN image, but it provides similar results in visibility domain to the CLEAN model (see Appendix \ref{appendix:images_comparison}). We consider that the SpM image better reconstructs the disk surface brightness distribution while the CLEAN model reconstructs an image with a sum of a number of point sources as shown in Figure \ref{fig:cont_images} (d), which do not reflect the disk structures precisely. Therefore, in the following sections, we adopt the SpM image to derive the physical properties of T Tau system.

\subsection{Dust Emissions from T Tau Sa and Sb}\label{sec:disk_ttaus}

As described in Section \ref{sec:spmimage}, two separate emissions were found around T Tau Sa and Sb. Figure \ref{fig:position_TTau} presents a close-up view of the Tau Sa and Sb regions. Two-dimensional Gaussian fitting was applied to each of them using the $\tt CASA$ task $\tt imfit$. The results are listed in Table \ref{tab:TTauS_disk}.

To confirm whether each of the two emissions originated from T Tau Sa or Sb, Figure \ref{fig:position_TTau} compares the image with stellar positions predicted by the stellar orbit model of the T Tau S binary in \cite{Kohler2016}, based on observational data spanning approximately 18 years. The coordinate systems of the binary were derived for the date of the ALMA observation (August 18, 2017). The offsets between the emission peaks and predicted stellar positions are calculated to be 9.3 mas (1.3 au) and 14.6 mas (2.1 au) for Sa and Sb, respectively, and will be roughly within errors involved in the calculations; e.g., the uncertainties of the model, a few mas \citep[][priv. comm]{Kohler2016} and $1\sigma$ positional errors for T Tau Sa and Sb in the SpM image, $\sim$1 and 5 mas, respectively. In addition, each total flux roughly fits each spectral energy distribution (SED) predicted by an accretion disk model \citep{Ratzka2009}. Hence, it can explain that the two emissions originate from T Tau Sa or Sb.

As shown in Table \ref{tab:TTauS_disk}, the best-fit sizes (i.e., the FWHM of the semi-major/semi-minor axes from the Gaussian fitting) of T Tau Sa and Sb were found to be 6$\times$4 au and 7$\times$3 au, respectively. These disk sizes are slightly larger than the effective spatial resolution of the SpM image ($\theta_{\rm eff} = 5\times 4$ au) and not resolved sufficiently. Hence, these sizes should be considered to be the conservative upper limits. The total flux density of T Tau Sb is a factor of seven smaller than T Tau Sa, which is in good agreement with a factor of eight given in \cite{Beck2020}. This implies that the actual disk size of T Tau Sb would be about three times smaller than that of T Tau Sa when we consider the scaling relation between the mm-continuum disk radii $R_{\rm mm}$ and luminosities $L_{\rm mm}$; $L_{\rm mm} \propto R_{\rm mm}^{2}$ \citep[][]{Tripathi2017,Andrews2018b, Hendler2020}.

\subsection{Disk Structure of T Tau N} \label{sec:disk_ttaun}
Here, we investigate the global disk properties of T Tau N derived from the SpM image and compare them with previous studies based on mid-infrared and millimeter observations. The T Tau N disk is known to be viewed as nearly face-on \citep{Akeson1998, Ratzka2009, Long2019, Manara2019, Beck2020}. We derived the inclination and PA of the disk on the image by fitting an ellipse to the outer ring, as described in Appendix \ref{appendix:disk_geometry}. As shown in Table \ref{tab:TTauN_disk}, the measured inclination of $25.2\pm1.1^{\circ}$ agrees well with $< 30^{\circ}$ derived from mid-infrared interferometric observations with very large telescope interferometer (VLTI) and SED simulations \citep{Ratzka2009} as well as with $\simeq28^{\circ}$ from visibility fitting using the same ALMA data \citep{Manara2019,Beck2020}.

Next, we derive a disk radius $r_{\rm d}$ using a curve-of-growth method similar to that described in \cite{Ansdell2016}. The disk radius is measured with successively larger photometric apertures on a deprojected image until the measured flux reaches $95\%$ of the total flux. As a result, $r_{\rm d }$ was calculated to be $24\pm 4$ au. The error on $r_{\rm d}$ is calculated by taking the range of radii within the uncertainties of the $95\%$ flux measurement. The obtained effective radius is in good agreement with the one ($r_{d} \simeq 21$ au) from \cite{Manara2019} in the same definition of the measurement. Most populated Taurus disks (spectral type earlier than M3) are known to be faint and compact \citep[Total flux of $< 100$ mJy at 1.3 mm, dust radii of $<40$ au;][]{Long2019}. Thus, the T Tau N disk can be regarded as a bright compact disk.

%-----------------------------Table Start-----------------------------

\begin{table*}[ht]
\label{table:ttau_propaties}
\caption{Properties of dust disks in T Tau system.}
\begin{tabularx}{\linewidth}{lrrrrc}
\toprule
         & \multicolumn{1}{l}{CLEAN} & \multicolumn{3}{r}{Sparse Modeling (SpM)}\\
           \cmidrule(lr){2-3}         \cmidrule(lr){4-6}
Source   & $F_{\nu} (>5\sigma)$ & Peak $I_{\nu}$ & $F_{\nu} (>I_{\rm DT})$ & Peak $I_{\nu}$ & Peak $I_{\nu}$ Position \\ 
         & (mJy)     & ($ \rm Jy~asec^{-2}$)     & (mJy)                  & ($\rm Jy~asec^{-2}$) & (RA, Dec)  \\ \hline
T Tau N  & 175.0       &    4.0                  & 174.4 & 9.1 & $(04^{h}21^{m}59^{s}.4475$, $+19^{d}32^{m}06^{s}.1731)$ \\
T Tau S (Sa+Sb) &  8.0 &    0.4                  &  7.9  &  $-$   &    $-$ \\
T Tau Sa &       $-$   &  $-$                    & 7.1   & 5.1 & $(04^{h}21^{m}59^{s}.4362$, $+19^{d}.32^{m}.05^{s}.5131)$ \\
T Tau Sb &       $-$   &  $-$                    & 0.8   & 0.8 & $(04^{h}21^{m}59^{s}.4365$, $+19^{d}.32^{m}.05^{s}.6131)$\\
\bottomrule
\end{tabularx}
\tablecomments{The first column lists the total flux ($F_{\nu}$) above $5\sigma (= 13~\rm mJy~asec^{-2})$ level and peak intensity (Peak $I_{\nu}$) on the CLEAN image. The last three columns list the total flux ($F_{\nu}$) above the $I_{\rm DT} (= 272~\rm mJy~asec^{-2})$ level , peak intensity (Peak $I_{\nu}$), and its position (RA, Dec) on the SpM image.}
\end{table*}
%-----------------------------Table End ------------------------------

%-----------------------------Table Start-----------------------------
\begin{table*}[ht]
\label{tab:TTauS_disk}
\caption{Results of 2D Gaussian Fits to T Tau Sa and Sb.}
\begin{tabularx}{\linewidth}{lccccccccc}
\toprule
Source  & $\theta_{\rm maj}$ & $\theta_{\rm maj}$ & $\theta_{\rm min}$ & $\theta_{\rm min}$ & PA & inclination & Peak $I_{\nu}$ & $F_{\nu}$  \\ 
       &  (mas) & (au) & (mas) & (au) & ($^\circ$) &  ($^\circ$) & (Jy/asec$^{2}$) & (mJy) \\ \hline
T Tau Sa  & 44.7 $\pm$ 0.7 & 6.4 $\pm$ 0.1 & 27.1 $\pm$ 0.4 & 3.9 $\pm$ 0.1 & 25.2 $\pm$ 1.1 & 52.8 $\pm$ 0.6 & 5.6 $\pm$ 0.1 & 7.7 $\pm$ 0.1 \\
T Tau Sb  & 49.7 $\pm$ 5.0 & 7.2 $\pm$ 0.7  &22.4 $\pm$ 2.3 & 3.2 $\pm$ 0.3 & 64.9 $\pm$ 4.6 &  63.2 $\pm$ 2.9 & 0.9 $\pm$  0.1 &  1.1 $\pm$ 0.1 \\
\bottomrule
\end{tabularx}
\tablecomments{Disk properties for T Tau Sa and Sb obtained using the $\tt imfit$ task in $\tt CASA$ to fit a 2D Gaussian on the SpM image. The task returns the total flux density ($F_{\nu}$) of the source along with the statistical uncertainty, the full width at half maximum (FWHM) along the semi-major ($ \theta_{\rm maj}$) and semi-minor ($\theta_{\rm min}$) axes, and the position angle (PA). The inclination is derived from $\theta_{maj}$ to $\theta_{min}$ ratio, assuming a perfect circle disk. Estimates of the uncertainties derived from $\tt imfit$ are based on \cite{Condon1997} and that of the inclination is derived from error propagation. Note that $\theta_{\rm maj}$ and $\theta_{\rm min}$ give upper limits, and these uncertainties in astronomical units do not account for the uncertainty in the distance to the source.}
\label{table:ttaus_model_fit}
\end{table*}
%------------------------Table End ------------------------------

%-----------------------------Table Start-----------------------------
\begin{table}[ht]
\label{tab:TTauN_disk}
\caption{Physical properties of T Tau N disk.}
\begin{tabularx}{\linewidth}{lrrr}
\toprule
Parameters  & Measurements  \\ \hline
position angle                         & $91.4\pm3.0~(^{\circ})$ \\
inclination                            & $25.2\pm1.1~(^{\circ})$ \\
disk radius: $r_{\rm d}$               & $166\pm 25$ (mas), $24\pm 4$ (au)\\
outer ring peak: $r_{\rm peak}$        & $109\pm 1$  (mas), $15.7\pm 0.1$ (au)\\
gap location: $r_{\rm gap}$            & $81 \pm 2$  (mas), $11.6 \pm 0.3$ (au) \\
gap width: $\Delta_{\rm I}$            & $0.28\pm 0.02$\\
gap depth: $\delta_{\rm I}$            & $1.22\pm 0.06$\\
\bottomrule
\end{tabularx}
\tablecomments{The physical parameters of the T Tau N disk are calculated from the SpM image and the radial intensity profile. These uncertainties in astronomical units do not account for the uncertainty in the distance to the source.}
\label{table:distance_ttau}
\end{table}
%-----------------------------Table End ------------------------------

%-----------------------------Start Figure------------------------------
\begin{figure}[t]
\centering
\includegraphics[width=0.45 \textwidth]{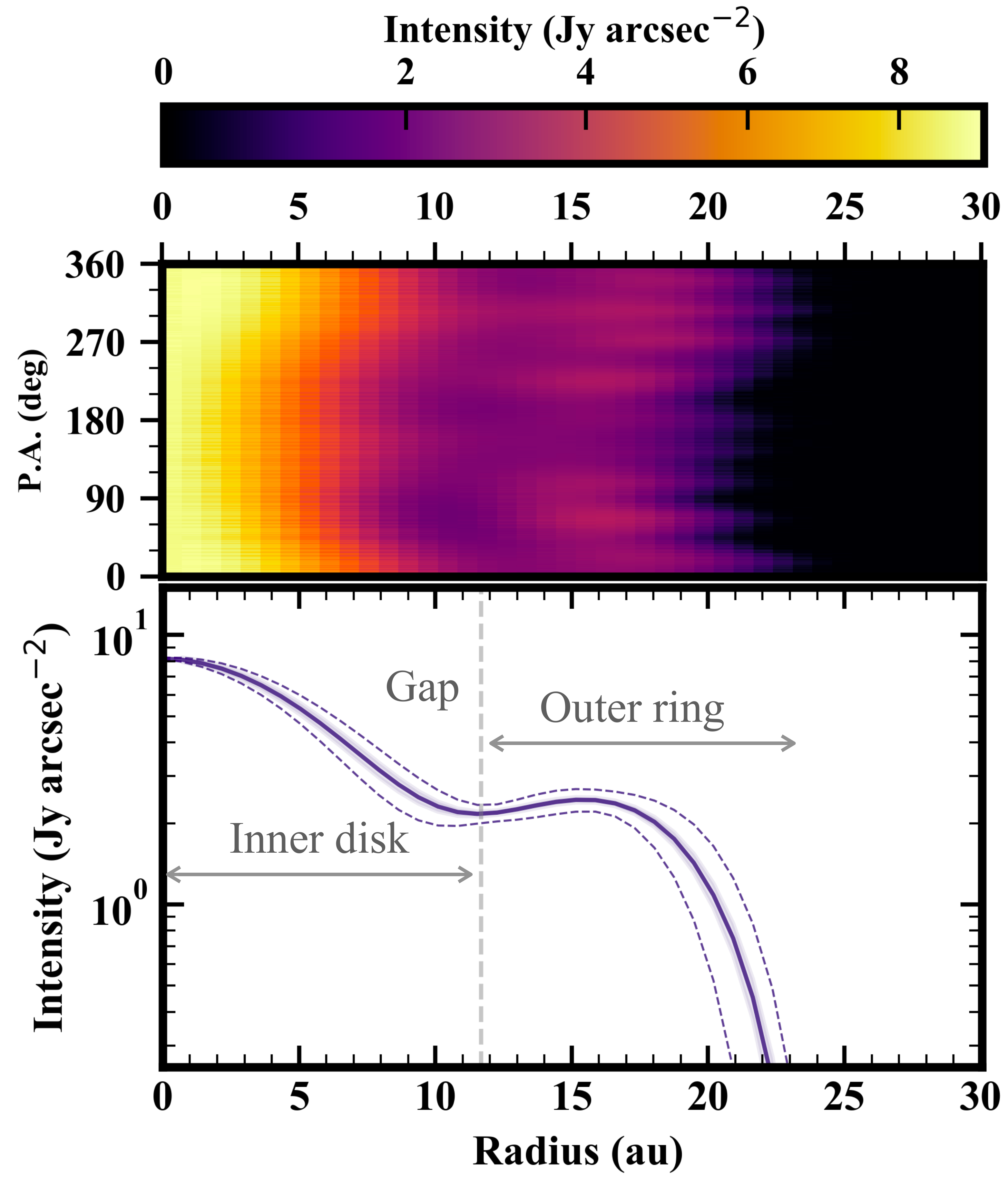}
\caption{Top panel: SpM image of T Tau N, which is deprojected into a map in polar coordinates to more clearly view the disk substructure. Bottom panel: Radial profiles of the intensity averaged over the full azimuthal angle on a logarithmic scale. The profile is linearly interpolated onto radial grid points spaced by 0.1 au with $\tt interpolate.interp1d$ in the $\tt SciPy$ module. The light purple ribbon shows the error of the mean at each radius, while the purple dashed lines show the standard deviation for comparison. For reference, the vertical dashed line marks the position of the gap, which correspond to the distance of 12 au.}
\label{fig:radialprofile}
\end{figure}
%-----------------------------End Figure---------------------------------

%-----------------------------Start Figure------------------------------
\begin{figure}[t]
\centering
\includegraphics[width=0.45 \textwidth]{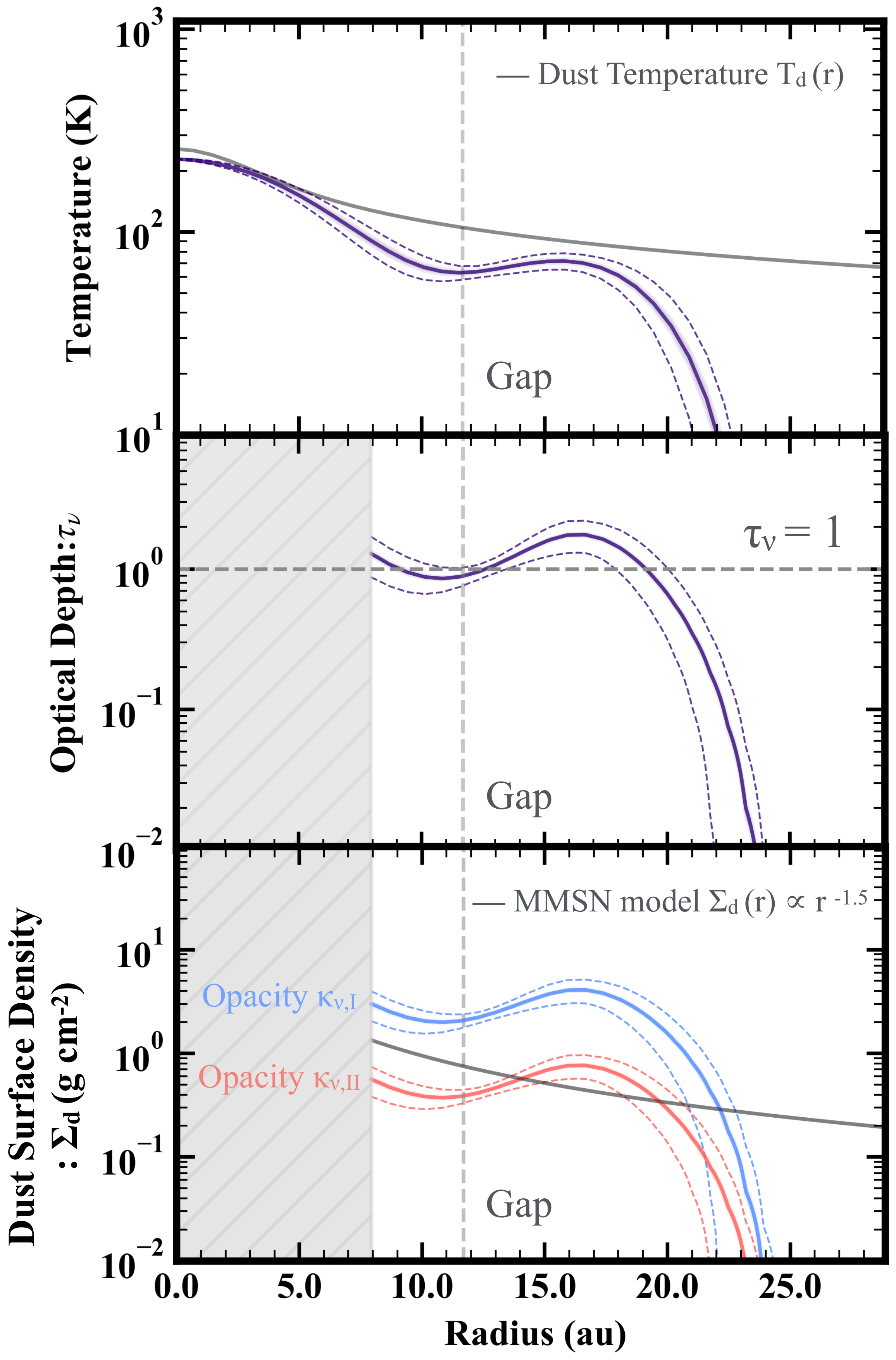}
\caption{Top panel: the radial profiles of the brightness temperature $T_{\rm br}(r)$ (purple line). The disk temperature model $T_{\rm d}(r)$ denotes the gray line. Middle panel: optical depth $\tau_{\nu}$(r) (purple line). The horizontal dashed gray line in the middle panel shows the threshold for the optical depth $\tau_{\nu} = 1$. Bottom panel: the dust surface densities $\Sigma_{\rm d}(r)$ for the two dust opacity models, compared to the \cite{Hayashi1981} dust surface density model for the MMSN (gray line). In the middle and bottom panels, the interior radii of 8 au are shaded because the optical thickness steepens by more than several orders of magnitude ($\tau_{\nu} \gg 1$) at the threshold ($r = 8$ au). The light colored ribbons show the error of the mean at each radius, while the dashed lines show the standard deviation for comparison. For reference, the vertical gray dashed lines indicate the annular gap location ($r = 12$ au) obtained from the radial intensity profile.}
\label{fig:physical_profile}
\end{figure}
%-----------------------------End Figure---------------------------------

\subsection{Gap structure in the T Tau N disk}\label{sec:intensityprifile_TTauN}
Figure \ref{fig:radialprofile} shows the deprojected and azimuthally averaged radial intensity profile $I_{\nu}(r)$ where $r= 0$ au is set to the peak intensity of the T Tau N disk. The uncertainty of the radial profile is evaluated as the error of the mean at each radius where we consider the effective spatial resolution of the major axis ($\theta_{\rm eff, maj}$) as the smallest independent unit. That is, the error is the standard deviation of each elliptical bin divided by the square root of the number of $\theta_{\rm eff, maj}$ spanning the whole azimuthal angle at each radial bin. For comparison, a standard deviation at each radius is also potted in the radial profile in Figure \ref{fig:radialprofile}.

We identify an annular gap (local minimum in $I_{\nu}(r)$ at $r_{\rm gap} = 11.6\pm 0.3$ au) and an outer peak (the local maximum in $I_{\nu}(r)$ at $r_{\rm peak} = 15.7\pm 0.1$ au) as well as in the radial intensity profile. We then adopt the same approach as in \cite{zhang2018} to measure the gap depth $\delta_{\rm I}$ and the gap width $\Delta_{\rm I}$. The gap depth is defined as $\delta_{\rm I} = I_{\nu}\left(r_{\text{peak}}\right) / I_{\nu}\left(r_{\text {gap}}\right)$. The gap width is defined as $\Delta_{\rm I}=\left(r_{\text {out}}-r_{\text {in}}\right) / r_{\text {out}}$, where $r_{\text {out}}$ and $r_{\text {in}}$ are the inner edges of the outer ring and the outer edge of the inner disk, respectively. The relationship between $I_{\rm edge} \equiv 0.5 \left \{I_{\nu}\left(r_{\rm peak}\right)+I_{\nu}\left(r_{\rm gap}\right) \right\}$ defines the edge locations. The edge location $r_{\text {in}}$ is defined as the smallest value $r$ satisfying the criteria $I_{\rm edge}= I_{\nu}(r_{\rm in }) $ and $ r < r_{\rm gap}$. Another edge location $r_{\rm out}$ is defined as the largest value satisfying the criteria $I_{\rm edge}= I_{\nu}(r_{\rm out}) $ and $ r_{\rm gap} < r < r_{\rm peak}$. The measured parameters are listed in Table \ref{tab:TTauN_disk}. In Section \ref{sec:planetmass}, the measured $\Delta_{\rm I} (= 0.28\pm 0.02)$ and $\delta_{\rm I} (= 1.22 \pm 0.06)$ are used to estimate the planetary mass under the hypothesis of planet-induced gap.

\subsection{Physical Properties of T Tau N Disk}\label{sec:physical_prop_ttaun}
Here, we derive the disk temperature $T_{\rm d}(r)$, optical depth $\tau_{\nu}(r)$, and dust surface density $\Sigma_{d}(r)$ of the T Tau N disk based on the SpM image to characterize the disk and substructure. We employ the radiative transfer equation expressed as
\begin{eqnarray}\label{eq:RT_equation}
I_{\nu}(r)=B_{\nu}\left(T_{d}(r)\right)\left(1-e^{-\tau_{\nu}}\right),
\end{eqnarray}
\noindent
where $B_{\nu}(r)$ and $T_{\rm d}(r)$ denote the full Planck function, and the dust temperature, respectively, and  $\tau_{\nu}(r)$ is the optical depth expressed as $\tau_{\nu}(r) = \kappa_{\nu} \Sigma_{\rm d}(r)$. Here, $\kappa_{\nu}$ and $\Sigma_{d}(r)$ denote the absorption dust opacity and the dust surface density, respectively. The brightness temperature $T_{\mathrm{br}}(r)$ can be calculated from Equation \ref{eq:RT_equation} such as
\begin{eqnarray}\label{tempreture}
T_{\mathrm{br}}(r)=\frac{h \nu}{k}\left[\ln \left(\frac{2 h \nu^{3}}{c^{2} I_{\mathrm{\nu}}(r) }+1\right)\right]^{-1},
\end{eqnarray}
\noindent
where $c$, $h$, and $k$ denote the speed of light, the Planck's constant, and the Boltzmann constant, respectively. Figure \ref{fig:physical_profile} (top panel) shows that $T_{\mathrm{br}}(r)$ reaches $257\pm 1$ K at the peak, and the average $T_{\mathrm{br}}(r)$ over the disk ($=\int T_{\rm br}(r)r dr / \int r dr$, where $r \leq 24$ au) is calculated to be $97\pm 1$ K. The average $T_{\mathrm{br}}(r)$ predominantly exceeds that predicted from the dust temperature model ($T_{d}\simeq 30-40$ K) which are simply scaled using the stellar luminosity \citep{Andrews2013, vanderPlas2016}. 

We should point out that the peak $T_{\rm br}$ is much higher than standard peak values ($\sim 20-100$ K) of other PPDs in the same observational wavelength and similar resolutions \citep[see Fig.4 in][]{Facchini2019}. Moreover, the average spectral index over the disk is estimated to be $\alpha_{\rm mm} = 1.9 \pm 0.1$ (see Appendix.\ref{appendix:spectral_index_TTauN}).

From the high brightness temperature and the low spectral index described above, the disk tends to be optically thick overall ($\tau_{\nu} \geq 1$), and the measured $T_{\rm br}(r)$ should represent the temperature of the emitting layer from the disk atmosphere. The innermost region ($r < 5$ au) seems to be thicker than the outer ring, and the brightness temperature should be close to the dust temperature near the disk surface in such a case. Therefore, we assume that $T_{\rm d}(r)$ is equal to $T_{\rm br}(r)$ at the optically thick region with $\tau_{\nu} \geq 1$. The disk temperature profile can be obtained as  $T_{\rm d}(r) = 360~(r/1~\rm au)^{-0.5}$ [K] by assuming $T_{\rm d}(r)$ has a power-law form, such as $T_{\rm d}(r) \propto r^{-0.5}$ \citep{Kenyon1987}. In this fitting, $T_{\rm d}(r)$ is smoothed with  $\theta_{\rm eff}$ to match the $T_{\rm br}(r)$ profile. It should be noted that a disk midplane temperature will generally be lower than $T_{\mathrm{br}}(r)$ at optically thick regions. Thus, dust surface densities (and dust masses) estimated in what follows would be lower limits in such a case.

We have estimated $\tau_{\nu}(r)$ and $\Sigma_{d}(r)$ by adopting $T_{\rm d}(r)$ derived above as the disk temperature. The optical depth $\tau_{\nu}(r)$ is calculated using the radiative transfer calculation of Equation \ref{eq:RT_equation} as:
\begin{eqnarray}\label{optical_depth}
\tau_{\nu}(r)=-\ln \left(1-\frac{I_{\nu}(r)}{B_{\nu}(T_{\rm d}(r))}\right).
\end{eqnarray}
\noindent
Figure \ref{fig:physical_profile} shows the derived $\tau_{\nu}(r)$ profile. $\Sigma_{d}(r)$ is also expressed as:

\begin{eqnarray}\label{dsut_surface_density}\label{eq:dust_surface_density}
\Sigma_{d}(r)=\frac{\tau_{\nu}(r)}{\kappa_{\nu}}.
\end{eqnarray}

\noindent
If we fix the disk temperature, another uncertainty in $\Sigma_{d}(r)$ comes from assumption of  the dust opacity $\kappa_{\nu}$, which usually depends on the grain size and many other factors. Here, we consider two independent dust opacity models (but keep not claiming which opacity model reproduces a ``better'' nature of the T Tau N). One is DSHARP opacity model $\kappa_{\nu,\rm I}$ \citep[$= 0.43~\rm cm^{2}~g^{-1}$;][]{Birnstiel2018} assuming a maximum grain size of 0.1 mm supported by recent (sub)mm polarization measurements of other Class II PPDs in the Taurus region \citep[][]{Bacciotti2018}. The model value is constrained by dust size distribution with reference to its measurements from (sub)mm observations. Another is a conventional model $\kappa_{\nu,\rm II}$ \citep[$= 2.3~\rm cm^{2}~g^{-1}$;][]{beckwith1991}, which can be expressed as $\kappa_{\nu} = 2.3 (\nu/230~\rm GHz)^{0.4} [cm^{2}~g^{-1}]$ and being simply parameterized because of the large uncertainties in the opacity. $\kappa_{\nu,\rm II}$  has been widely used for PPDs \citep[e.g.,][]{Williams2011} and being supported by spatially resolved multi-wavelength continuum observations of other PPD \citep{Lin2021}.

The final results of $\Sigma_{d}(r)$ using $\kappa_{\nu,\rm I}$ and $\kappa_{\nu,\rm II}$ are plotted in Figure \ref{fig:physical_profile} together with that of the minimum mass solar nebula \citep[MMSN; $\Sigma_{d}(r) = 30~(r/1~\mathrm{au})^{-1.5}~\mathrm{[g~cm^{-2}]}$;][]{Weidenschilling1977, Hayashi1981}. We found that the dust surface density profiles of the T Tau N disk are locally more massive than the MMSN by a factor of 15 for $\kappa_{\nu,\rm I}$ and 3 for $\kappa_{\nu,\rm II}$ around the outer ring, but it sharply decreases at the disk edge. For comparison, the dust surface density profiles of the disks in Ophiuchus, Taurus-Auriga \citep{Andrews2015PASP}, and Lupus \citep{Tazzari2017} generally appear less massive than the MMSN, while only a few of them have a comparable or larger mass. The T Tau N disk, despite being a small dust disk, would be regarded to be more massive than typical disks in the low-mass star-forming regions.

The mm-dust mass $M_{\rm dust}$ of the outer ring ($r_{\rm gap} \leq r \leq r_{\rm d}$) can be computed using the obtained $\Sigma_{d}(r)$, which is defined as $M_{\mathrm{dust}}=\int_{r_{\rm gap}}^{r_{d}} \Sigma_{d}(r) 2 \pi r dr$. The dust ring mass results in a wide range of values depending on the dust opacity; $\sim 104~M_{\oplus}$ for $\kappa_{\nu, \rm I}$ and $\sim 20~M_{\oplus}$ for $\kappa_{\nu, \rm II}$. We note that these dust ring masses should be considered as lower limit due to the uncertainty of midplane temperature. In the range of the inferred dust masses, the outer ring of T Tau N is roughly as massive as the outer ring (at a location of $\sim100$ au, $M_{\rm dust}\sim 67~M_{\oplus}$) in Herbig Ae star MWC 480, as located in the Taurus region \citep{Liu2019}.

To check whether the T Tau N disk is gravitationally stable, Toomre $Q$ \citep{Toomre1964} was calculated using the formula; $Q \equiv c_{s} \Omega_{\mathrm{K}}/\pi G \Sigma_{\mathrm{gas}}$, where $c_{s}$ is sound speed, $\Omega_{\mathrm{K}}$ is angular velocity, $G$ is the gravitational constant, and $\Sigma_{\mathrm{gas}}$ is gas surface density. If the disk follows the criterion $Q \lesssim 1.5$, the disk is gravitationally unstable and grows spiral arms \citep[][]{Laughlin1994}. Here, we employed the Toomre $Q$ under the standard assumption of $\Sigma_{\mathrm{gas}}/\Sigma_{\mathrm{dust}} = 100$ \citep{Bohlin1978}. In both $\kappa_{\nu, \rm I}$ and $\kappa_{\nu, \rm II}$, the Toomre Q values exceed the unity around the outer ring area; $Q \gtrsim 3$ for $\kappa_{\nu, \rm I}$ and $Q \gtrsim 10$ for $\kappa_{\nu, \rm II}$. The T Tau N disk thus appears to be gravitationally stable. It should be noted that a secular gravitational instability (requiring high gas-to-dust ratios $<100$ and low viscous parameter $\alpha \lesssim 10^{-3}$) can generate a ring-like structure in the disk \citep{Takahashi2014,Takahashi2016}. That being said, the required physical parameters remain highly uncertain at this stage, and it cannot conclude the possibility of secular gravitational instability as the origin of the rings yet.

\section{Discussion}\label{sec:discussion}
\subsection{Origins of Gap in the T Tau N Disk}

Recent high-resolution observations have revealed multiple annular gap structures in bright giant disks \citep[e.g.,][]{Andrews2018DSHARP, cieza2020}. Several detections of gaps or cavities in compact disks have also been reported so far; the transitional disks around XZ Tau B \citep[disk radius of 3 au, cavity radius of 1.3 au,][]{Osorio2016} and around several candidates \citep[see][for details]{Pinilla2018}, and the annular gap structure in the disks around a single star SR 4 \citep[$r_{\rm gap} = 11$ au, $r_{\rm d} = 31$ au;][]{huang2018}, DoAr 33 \citep[$r_{\rm gap} = 9$ au, $r_{\rm d} = 27$ au;][]{huang2018}, WSB 52 \citep[$r_{\rm gap} = 21$ au, $r_{\rm d} = 32$ au;][]{huang2018}, CIDA 1 \citep[$r_{\rm gap} = 8$ au, $r_{\rm d} =40$ au;][]{Pinilla2021}, J0433 \citep[$r_{\rm gap} = 15$ au, $r_{\rm d} = 46$ au;][]{Kurtovic2021}, and one of a binary system GQ Lup A \citep[$r_{\rm gap} = 8$  au, $r_{\rm d} = 20$ au;][]{Long2020}. The T Tau N case is very similar to SR 4, DoAr 33, and GQ Lup A in terms of the radius of the disk and gap location. In previous studies on gap origins in disks \citep[e.g.,][]{huang2018, Long2018}, two main possibilities have been investigated, that is, snow line and planet origins.

The snow line, which is also referred to as an ice sublimation front, is the location in the disk midplane where dust opacity and collisional growth are expected to change, producing features such as ring-like substructures seen in continuum images \citep[][]{zhang2015, Okuzumi2016}. An estimate of the snowline location inferred from disk midplane temperature models \citep[e.g.,][]{Dullemond2001} should be calculated and is confirmed by comparing the gap location. However, the brightness temperature of the T Tau N disk is much higher than that of the regular disk, and this disk appears optically thick at 1.3 mm. It can be thus challenging to find a reasonable disk midplane temperature model that matches observations. This problem would be solved by observing the optically thin disk at lower wavelengths to determine an adequate model.

\subsection{Planetary Origin and Planet Mass Estimates}\label{sec:planetmass}

%-----------------------------Start Figure------------------------------
\begin{figure}[t]
\centering
\includegraphics[width=0.5 \textwidth]{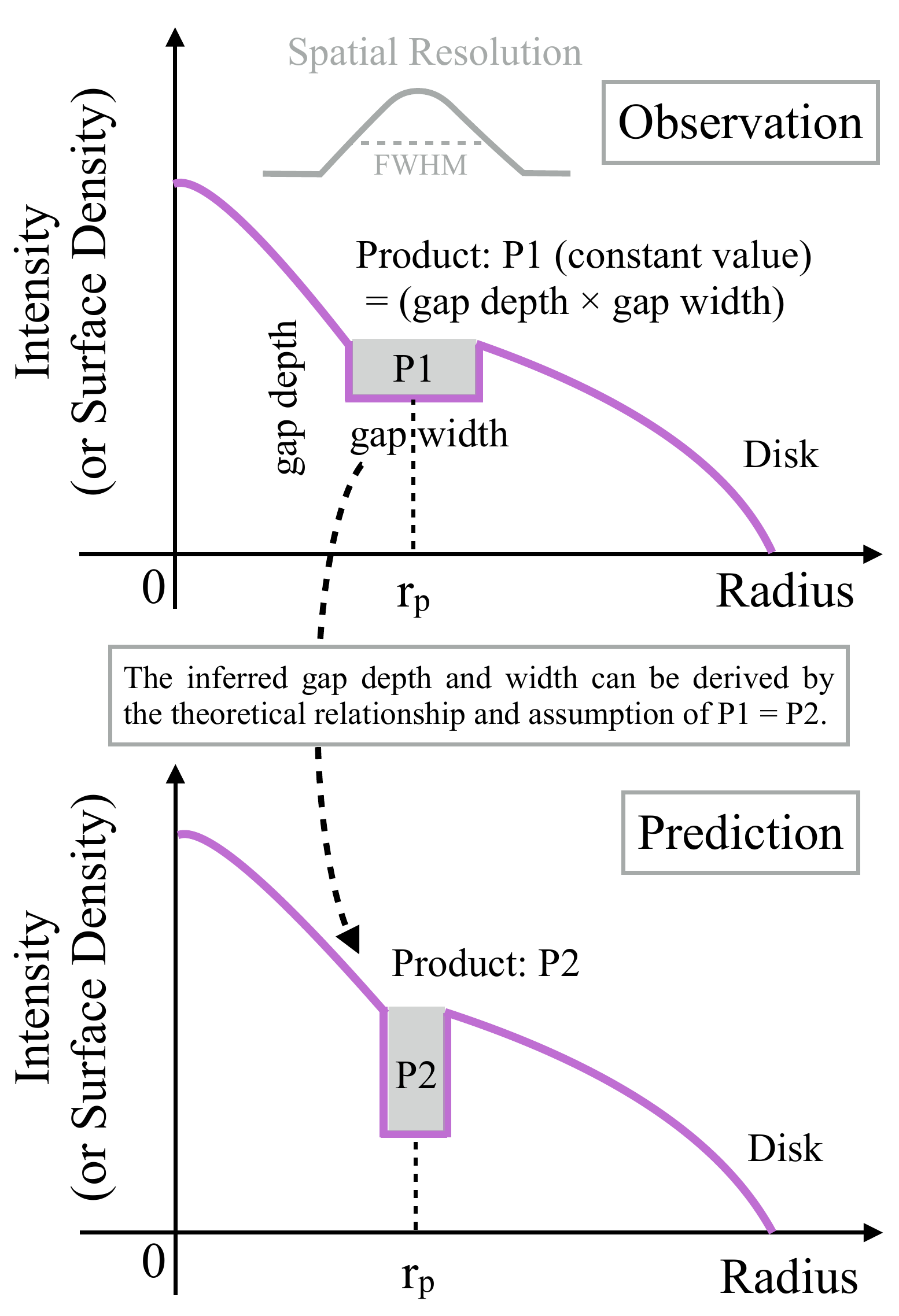}
\caption{Schematic view of the conservation relation with the gap width and depth in radial intensity (or surface density) profile for the T Tau N disk.}
\label{fig:gap_estimate}
\end{figure}
%-----------------------------End Figure---------------------------------

%-----------------------------Start Figure------------------------------
\begin{figure*}[t]
\centering
\includegraphics[width=0.95 \textwidth]{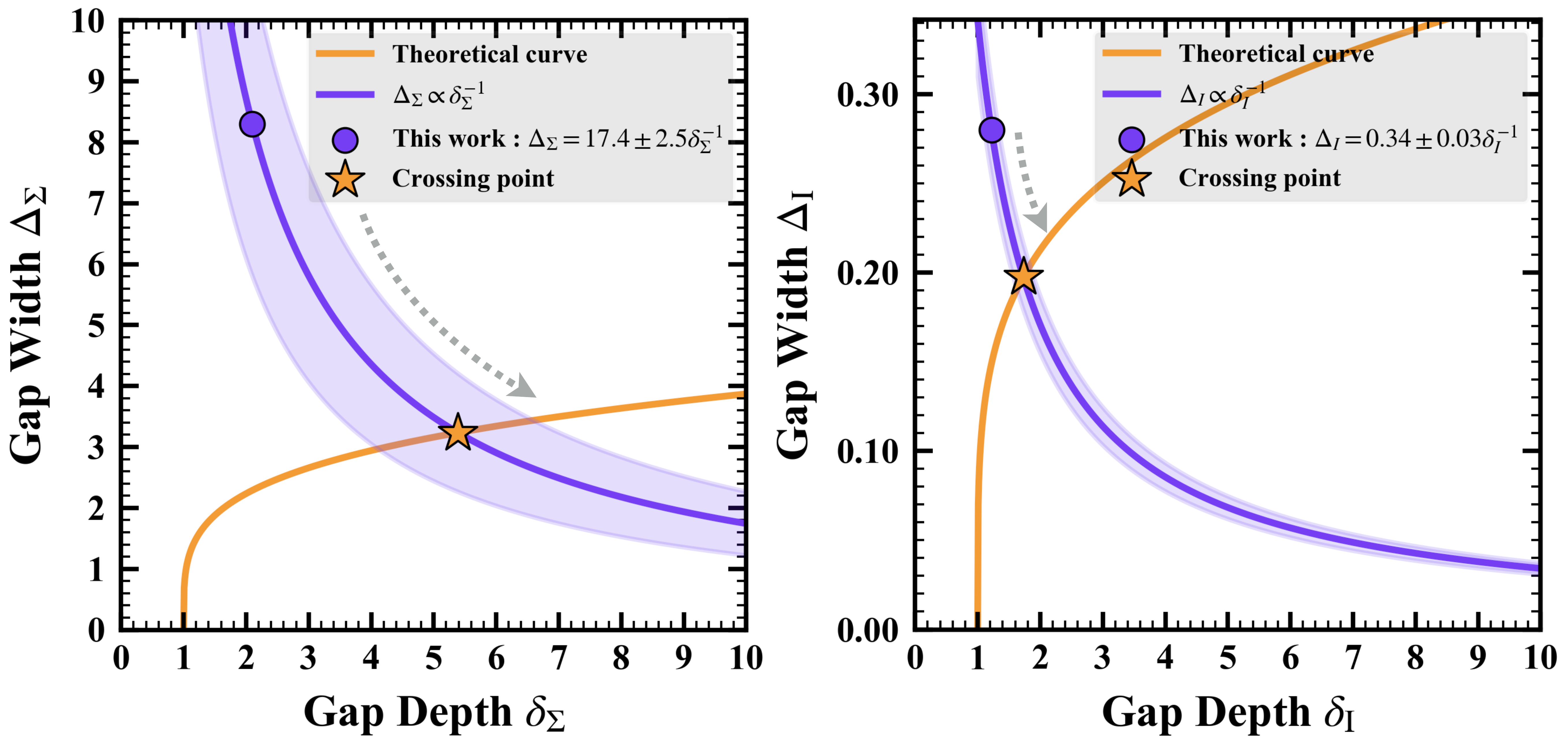}
\caption{Relationships between the gap width and depth controlled by the planetary mass using Eq.\ref{eq:kanagawa_theory} (orange curve in the left panel) and Eq.\ref{eq:zhang_theory} (orange curve in the right panel). In each panel, the blue circle denotes the measured values obtained from the observation, the blue curve denotes the relationship that the gap depth times the gap width retains the values $(\Delta_{\Sigma} \propto \delta^{-1}_{\Sigma}$, or $\Delta_{\rm I} \propto \delta^{-1}_{\rm I})$ derived from the observation, and the star mark shows the cross point between the blue curve and the orange curve indicating the prediction of a set of the gap width and depth. The light blue ribbons show the uncertainties of the gap width due to $1\sigma$ errors in both the measured gap depth and width.}
\label{fig:planet_mass}
\end{figure*}
%-----------------------------End Figure---------------------------------

Another possible origin of this gap is the planet clearing of the disk material. Below, we estimate planetary masses by applying two different methods that connect the planetary mass and gap shape \citep{Kanagawa2015,Kanagawa2016, zhang2018}.

We first apply the relationship between the planet mass and the width and depth of the gaseous gap according to the theory proposed by \cite{Kanagawa2015, Kanagawa2016}. The gap depth and width are defined as the difference between the initial and gap-formed surface density profiles in the theory, but this definition cannot be adopted in our case as the initial model cannot be set because of a lack of complete information of $\Sigma_{d}(r)$. Instead of the original definition, we adopt the gap width $\Delta_{\Sigma}$ and depth $\delta_{\Sigma}$ given in the dust surface density $\Sigma_{d}(r)$. As shown in Equation \ref{eq:dust_surface_density}, $\Sigma_{d}(r)$ is simply calculated by dividing the optical depth $\tau_{\nu}(r)$ by the constant opacity models ($\kappa_{\rm \nu, I}$ or $\kappa_{\rm \nu,II}$). Therefore, the gap width and depth of the dust surface density profile would not change regardless of which of the two opacities are used. Here, we apply a Gaussian fit to $\Sigma_{\rm d}(r)$ at the gap area ($r = 8-16$ au) in a similar manner with \cite{Cox2020}, by using the least-squares method implemented in $\tt optimize.leastsq$ from $\tt SciPy$ \citep{Jones2001}. The uncertainties are the statistical uncertainties from the Gaussian fitting. The FWHM gap width $\Delta_{\Sigma}$ is derived as $8.3 \pm 0.1$ au at a gap location of $r_{\Sigma,\rm gap} = 10.89 \pm 0.02$ au. We define the gap depth $\delta_{\Sigma} = \Sigma_{\rm peak}/\Sigma_{\rm gap}$, where $\Sigma_{\rm peak}$ is the local peak of the outer ring, and $\Sigma_{\rm gap}$ is the local minimum at $r_{\Sigma,\rm gap}$. The gap depth was calculated to be $\delta_{\Sigma} = 2.1 \pm 0.3$.

Here, we assume that the dust is well coupled to the gas content of the disk, and the radial location of the planet is at $r_{\rm\Sigma,gap}$. Note that our defined gap depth may be underestimated because it may be shallower than the original one. We use the relationship between the planetary mass $M_{\rm p}$ and the gap depth $\delta_{\Sigma}$ as follows \citep[Eq.7 in][]{Kanagawa2015}:
\begin{eqnarray}\label{eq:kanagawa_theory_gapdepth}
\frac{M_{\mathrm{p}}}{M_{*}}=0.16(\delta_{\Sigma}-1)^{0.5}\left(\frac{h_{\mathrm{\Sigma,gap}}}{r_{\mathrm{\Sigma,gap}}}\right)^{2.5}\left(\frac{\alpha_{\rm  vis}}{10^{-3}}\right)^{0.5},
\end{eqnarray}

\noindent
where $h_{\rm\Sigma,gap}$ is the scale height at $r_{\mathrm{\Sigma,gap}}$, and $\alpha_{\rm  vis}$ is the viscous parameter \citep{Shakura1973}. We also use the relation with the gap width $\Delta_{\mathrm{\Sigma}}$ as follows \citep[Eq.5 in][]{Kanagawa2016}:

\begin{eqnarray}\label{eq:kanagawa_theory_gapwidth}
\frac{M_{\mathrm{p}}}{M_{*}}=0.19 \left(\frac{\Delta_{\mathrm{\Sigma}}}{r_{\mathrm{\Sigma,gap}}}\right)^{2}\left(\frac{h_{\mathrm{\Sigma,gap}}} {r_{\mathrm{\Sigma,gap}}}\right)^{1.5}\left(\frac{\alpha_{\rm  vis}}{10^{-3}}\right)^{0.5}.
\end{eqnarray}

\noindent
By eliminating the planetary mass in the above two equations, Eq.\ref{eq:kanagawa_theory_gapdepth} and  Eq.\ref{eq:kanagawa_theory_gapwidth}, the relationship between $\delta_{\Sigma}$ and $\Delta_{\Sigma}$ can be obtained as follows:

\begin{eqnarray}\label{eq:kanagawa_theory}
\Delta_{\Sigma} = 0.92 (\delta_{\Sigma} -1)^{0.25} \left(\frac{r_{\rm\Sigma,gap}}{\rm 1~au}\right) \left(\frac{h_{\rm\Sigma,gap}}{r_{\rm\Sigma,gap}}\right)^{0.5} \rm au
\end{eqnarray}
\noindent
Using $T_{\rm d}(r)$, the aspect ratio $h_{\rm\Sigma,gap}/r_{\mathrm{\Sigma,gap}}$ was calculated to be 0.05, and the viscous parameter is set to be $10^{-3}$ for the T Tau N disk, as in \cite{Kanagawa2015}. We found that the derived depth and width are too shallow and too wide compared to the theoretical curve. \cite{Nomura2016} reported that owing to beam smearing, the derived measurements, $\delta_{\Sigma}$ and $\Delta_{\Sigma}$ give a lower limit and upper limit, respectively. Following the discussion by \cite{Nomura2016}, we assume that the gap depth times the gap width conserves the value derived from the observations (i.e., $\Delta_{\Sigma} \propto \delta^{-1}_{\Sigma}$) and $r_{\rm p}$ remains fixed, as illustrated in Figure \ref{fig:gap_estimate}. Thus, the relation can be plotted in the left panel of Figure \ref{fig:planet_mass}. The crossing point between the two curves is located at the width and depth of $\Delta_{\Sigma} = 3.2 \pm 0.3$ au and $\delta_{\Sigma} = 5.4 \pm 1.3$ under the condition of $\Delta_{\Sigma} = 17.4 \pm 2.5~\delta_{\Sigma}$ au. The crossing point and the use of Eq.\ref{eq:kanagawa_theory_gapdepth} and Eq.\ref{eq:kanagawa_theory_gapwidth} give the planetary mass of $1.4 \pm {0.2}$ $\rm M_{Saturn}$. The error range of the planetary mass results from the uncertainty of the product, which can change the location of the crossing point.  

Next, we consider another relationship in \cite{zhang2018}. This approach defines the gap depth $\delta_{\rm I}$ and width $\Delta_{\rm I}$ in $I_{\nu}(r)$ in Section \ref{sec:intensityprifile_TTauN} without assuming a functional form for the substructures or an initial surface density; it has the relationships to derive the planet mass from the measured $\delta_{\rm I}$ and $\Delta_{\rm I}$ in Section \ref{sec:intensityprifile_TTauN}. We now use the relationship between the planet mass and the gap depth $\delta_{\rm I}$ \citep[Eq.24 in][]{zhang2018}:

\begin{eqnarray}\label{eq:zhang_theory_gapdepth}
\frac{M_{\mathrm{p}}}{M_{*}}=0.073\left(\frac{\delta_{\rm I}-1}{C}\right)^{1 / D}\left(\frac{h_{\mathrm{gap}}} {r_{\mathrm{gap}}}\right)^{2.81}\left(\frac{\alpha_{\rm  vis}}{10^{-3}}\right)^{0.38}
\end{eqnarray}

\noindent
We also used the one with the gap width $\Delta_{\mathrm{\rm I}}$ \citep[Eq.22 in ][]{zhang2018}:

\begin{eqnarray}\label{eq:zhang_theory_gapwidth}
\frac{M_{\mathrm{p}}}{M_{*}}=0.115\left(\frac{\Delta_{\mathrm{I}}}{A}\right)^{1 / B}\left(\frac{h_{\mathrm{gap}}} {r_{\mathrm{gap}}}\right)^{0.18}\left(\frac{\alpha_{\rm  vis}}{10^{-3}}\right)^{0.31}
\end{eqnarray}

\noindent
By eliminating the planetary mass in the two equations, Eq.\ref{eq:zhang_theory_gapdepth} and  Eq.\ref{eq:zhang_theory_gapwidth}, the relationship between $\Delta_{\rm I}$ and $\delta_{\rm I}$ can be obtained as follows.
\begin{eqnarray}\label{eq:zhang_theory}
\Delta_{\mathrm{I}}=\mathrm{A}\left[0.635\left(\frac{\delta_{\mathrm{I}}-1}{\mathrm{C}}\right)^{1 / D} \right. \nonumber\\
\left. \times\left(\frac{h_{\mathrm{gap}}}{r_{\mathrm{gap}}}\right)^{2.63}\left(\frac{\alpha_{\mathrm{vis}}}{10^{-3}}\right)^{0.07}\right]^{\mathrm{B}}
\end{eqnarray}
\noindent
where $A$, $B$, $C$, and $D$ are constant parameters introduced by \cite{zhang2018}, and depend on the gas surface density $\Sigma_{\rm g}$ and the maximum grain size ($s_{\rm max}$ = $0.1\sim 10$ mm). Figure 18 of \cite{zhang2018} shows the relationship between a gas surface density and an averaged dust surface density $\Sigma_{\rm d}$ at an outer disk (or ring) for hydrodynamical simulations. We can then use their Figure 18 to estimate the gas surface density $\Sigma_{\rm g}$ based on the average $\Sigma_{\rm d} (= 3.4~\rm g~cm^{-2})$ at the outer ring of T Tau N and the aspect ratio $h_{\rm gap}/r_{\rm gap} (= 0.05$). Finally, the four parameters ($A = 1.11$, $B = 0.29$, $C = 0.0478$, and $D = 1.23$) are selected from Table 1 and 2 of \cite{zhang2018}, when considering the estimated $\Sigma_{\rm g} (> 100~\rm g~cm^{-2})$ and the maximum dust particle size \citep[$s_{\rm max} = 0.1$ mm;][]{Bacciotti2018} in the disk.

$\Delta_{\rm I}$ is calculated as a function of $\delta_{\rm I}$, as shown in Figure \ref{fig:planet_mass}, where we used the obtained parameters for ($A$, $B$, $C$, and $D$), the aspect ratio $(h_{\rm gap}/r_{\rm gap})$ of $0.05$, and the viscous parameter of $\alpha_{\rm vis} = 10^ {-3}$. A conservation relation derived from the measured $\Delta_{\rm I}$ and $\delta_{\rm I}$ can also be obtained by assuming that the product of the gap depth and width conserves on the radial intensity profile (i.e., $\Delta_{\rm I} \propto  \delta_{\rm I}^{-1}$), as shown in the right panel of Figure \ref{fig:planet_mass}. We obtain the relation $\Delta_{\rm I} = 0.34 \pm 0.03~\delta_{\rm I}$ for $\Delta_{\rm I} = 0.19 \pm 0.01$ and $\delta_{\rm I} = 1.78 \pm 0.11$. The crossing point between the two curves gives a planetary mass of $1.2 \pm 0.1~\rm M_{Saturn}$, which agrees well with $1.4 \pm 0.2~\rm M_{Saturn}$ derived from the analytic formula by \cite{Kanagawa2015, Kanagawa2016} within the uncertainties involved.

We assumed the viscous parameter, $\alpha_{\rm  vis} = 10^{-3}$ for the above estimate. If we take $\alpha_{\rm  vis}$ over a wide range of $\alpha_{\rm  vis} = 10^{-2} - 10^{-4}$, the derived planetary masses vary by a factor of $\sim 3$, and are calculated to be $0.5 - 4.5~M_{\rm Saturn}$ for the analytic formula by \cite{Kanagawa2015, Kanagawa2016} and $0.5 - 2.7~M_{\rm Saturn}$ for the analytic formula by \cite{zhang2018}. Even considering the wide range of $\alpha_{\rm vis}$, the planet mass is still similar to Saturn's mass. In addition, the gap location ($r = 12$ au) is close to Saturn's orbit, and T Tau N is an interesting example that is analogous to the solar planetary system. 

There are two other cases of gaps at $r \simeq 10$ au in the compact disks, indicating the presence of planets: SR 4 \citep{zhang2018} and GQ Lup A \citep{Long2020}. Both cases indicate upper limits of planet masses of $M_{p} \lesssim 7.2~M_{\rm Saturn}$ for SR 4 \citep[$F_{\nu} = 69$ mJy at 1.3 mm;][]{Andrews2018b} and $M_{p} \lesssim 0.1~M_{\rm Saturn}$ for GQ Lup A \citep[$F_{\nu} = 28$ mJy at 1.3 mm;][]{Wu2017}, by using the the same manner as \cite{zhang2018} taken from a gap width alone for $\alpha_{\rm  vis} = 10^{-3}$ and $s_{\rm max} = 0.1~\rm mm$. While there are a few samples of the inferred planet mass for the compact disks at this stage, it could be a correlation between planet mass and (sub)millimeter disk flux (or disk mass) in such disks, suggesting that more massive disks tend to produce more massive planets \citep{Lodato2019}. Planets inferred to be forming in the larger DSHARP disks are roughly in a Neptune mass group at the outer disk ($r>10$ au) and in a Saturn-Jupiter mass group at the inner disk ($r\simeq10$ au) \citep[see Fig.21 in ][]{zhang2018}, i.e., the planet mass could be higher at smaller radii. Thus, investigating further the tendency for compact disks versus large disks would lead to an intriguing study for understanding planet mass induced by disk size.

According to the core accretion model of giant planets in a minimal mass solar nebula, the optimal formation site is believed to be $r = 5-10$ au \citep{Helled2014}, although the distance at which a giant gas planet can form could be greater than 10 au on the assumption of pebble accretion \citep{Lambrechts2012} or the maximum rate of planetesimal accretion \citep{Rafikov2011}. Investigations of such an inner $5 - 10$ au region for compact disk sources, including our target with higher spatial resolution, is also of great interest, and could be very valuable for comparing theoretical and observational studies.

\section{Conclusions}
By using the super-resolution imaging with sparse modeling \citep{yamaguchi2020}, we investigated a young triple system T Tau using ALMA 1.3 mm archival data. A summary of our findings is as follows:

\begin{enumerate}
\item{The imaging drastically improves the spatial resolution on the continuum image of the T Tau system. We then find an annular emission gap in the T Tau N disk and two new emissions around T Tau Sa and Sb.}

\item{The effective spatial resolution of the image achieves $\sim30\%$ ($38\times27$ mas or $5\times4$ au) compared with the CLEAN beam size confirmed by tests evaluating the response to artificial point source injections.} This result is in good agreement with the prediction that interferometric imaging can use visibility amplitudes at maximum baselines for deriving source structures by $\sim 30\%$ (or 1/3) of the synthesized beam size.

\item{Each position of the separated two emissions around T Tau Sa and Sb is in good agreement within their uncertainties, with each one predicted by the stellar orbital model in \cite{Kohler2016}. In addition, each total flux roughly fits each SED predicted by an accretion disk model \citep{Ratzka2009}. The two emissions can thus be regarded as dust emissions originating from the circumstellar disks of T Tau Sa and Sb. The dust disk sizes of T Tau Sa and Sb are smaller than $6\times4$ au ($45\times27$ mas) and $7\times3$ au ($50\times22$ mas), respectively. The total flux density of T Tau Sb is about seven times lower than T Tau Sa. This ratio implies that the actual disk size of T Tau Sb would be smaller than that of T Tau Sa when considering general scaling relations between disk properties \citep{Tripathi2017, Andrews2018b, Hendler2020}.}

\item{The T Tau N disk has a radius of $24\pm 4$ au radius enclosing $95\%$ of the total flux and has an annular gap at $r = 11.6\pm0.3$ au. Its total flux is as large as $174$ mJy, comparable with that of much larger disks. The disk then shows the high brightness temperature of $257\pm1$ K at the peak and the low spectral index of $1.9\pm0.1$, suggesting the optically thick at 1.3 mm. The lower-limited dust surface density appears higher than the MMSN case locally at the outer ring, even though a majority of disks in the low mass star-forming regions generally appear less massive than the MMSN \citep{Andrews2015PASP, Tazzari2017}. The T Tau N disk, despite being a small dust disk, would be regarded to be more massive than regular disks. Meanwhile, given the relatively high values of Toomre $Q$ parameter ($Q > 3$) at the outer ring, it appears to be gravitationally stable.}

\item{We considered a possibility for the origin of the gap in the T Tau N disk by using two different methods that connect the planetary mass and gap shape. If we take a viscous parameter over a wide range of $10^{-2}-10^{4}$, the derived planetary masses are similar to Saturn's mass; $0.5 - 4.5~M_{\rm Saturn}$ for the analytic formula by \cite{Kanagawa2015, Kanagawa2016} and $0.5 - 2.7~M_{\rm Saturn}$ for the analytic formula by \cite{zhang2018}.}

\end{enumerate}

Our super-resolution imaging is potentially impressive, but the reliability of the resolution and the findings for the T Tau system would depend on the confirmation of the substructure by future observations with better angular resolution and sensitivity. Ultimately, ALMA observations with a higher spatial resolution comparable to the effective resolution, at least 30 mas, can assess the consistency of the super-resolution imaging and confirm the existence of the substructure.

\acknowledgements
We thank the anonymous referee for all of the comments and advice that helped improve the manuscript and contents of this study. We also thank all of the East Asian ALMA staff members at NAOJ for their kind support and Sai Jinshi for the helpful conversation. We are grateful to $\rm R.K\ddot{o}hler$ for providing us with the predicted coordinate data of the target source. We finally thank Editage (\url{https://www.editage.com}) for English language editing. MY was financially supported by the Public Trust Iwai Hisao Memorial Tokyo Scholarship Fund and the Sasakawa Scientific Research Grant from the Japan Science Society. This work was financially supported in part by Japan Society for the Promotion of Science (JSPS) KAKENHI grants No. 17H01103, 18H05441, 19K03932 (TM), and JP17K14244, JP20K04017 (TT), and 18H05441, 19K03910, 20H00182 (HN), and 20H01951 (SI), 18H05442, 15H02063, and 22000005 (MT). This study uses the following ALMA data: ADS/JAO.ALMA$\#2016.1.01164.\rm S.$ ALMA is a partnership of ESO (representing its member states), NSF (USA) and NINS (Japan), together with NRC (Canada), MOST and ASIAA (Taiwan), and KASI (Republic of Korea), in cooperation with the Republic of Chile. The Joint ALMA Observatory is operated by ESO, AUI/NRAO, and NAOJ." Data analysis was in part carried out on the multi-wavelength data analysis system operated by the Astronomy Data Center (ADC), National Astronomical Observatory of Japan. This study used data from the European Space Agency (ESA) mission $\it Gaia$ (\url{https://www.cosmos.esa.int/gaia}), processed by the {\it Gaia} Data Processing and Analysis Consortium (DPAC, \url{https://www.cosmos.esa.int/web/gaia/dpac/consortium}). Funding for the DPAC has been provided by national institutions, in particular the institutions participating in the $\it Gaia$ Multilateral Agreement.

\software{AnalysisUtilities (\url{https://casaguides.nrao.edu/index.php?title=Analysis_Utilities}), Astropy \citep{astropy2013}, CASA \citep{mcmullin2007}, DIFMAP \citep{Shepherd1994}, matplotlib \citep{hunter2007matplotlib}, PRIISM \citep{nakazato2020}, SciPy \citep{Jones2001}}, least-squares fitting \citep{hammel2020}

% APPENDIX ##############################################################
\restartappendixnumbering  
%  If the command \restartappendixnumbering is used anywhere in the manuscript the figure and table numbers will be reset to one in each section with the section letter appended to the front, e.g. Figure A1, A2 or Table B1 and C1. Equations labeling always follows this format.
\appendix

\section{Comparison between CLEAN and SpM Images}\label{appendix:images_comparison}
As shown in Fig.\ref{fig:cont_images}, we have three images for the T Tau system: the beam-convolved CLEAN image (Fig. \ref{fig:cont_images}(c)), the CLEAN components or CLEAN model (Fig. \ref{fig:cont_images}(d)), and SpM (Fig. \ref{fig:cont_images}(b)). Here, we compare the three images in visibility domain and discuss which image can be the best used for image analyses.

\begin{figure*}[t]
\centering
\includegraphics[width=0.95 \textwidth]{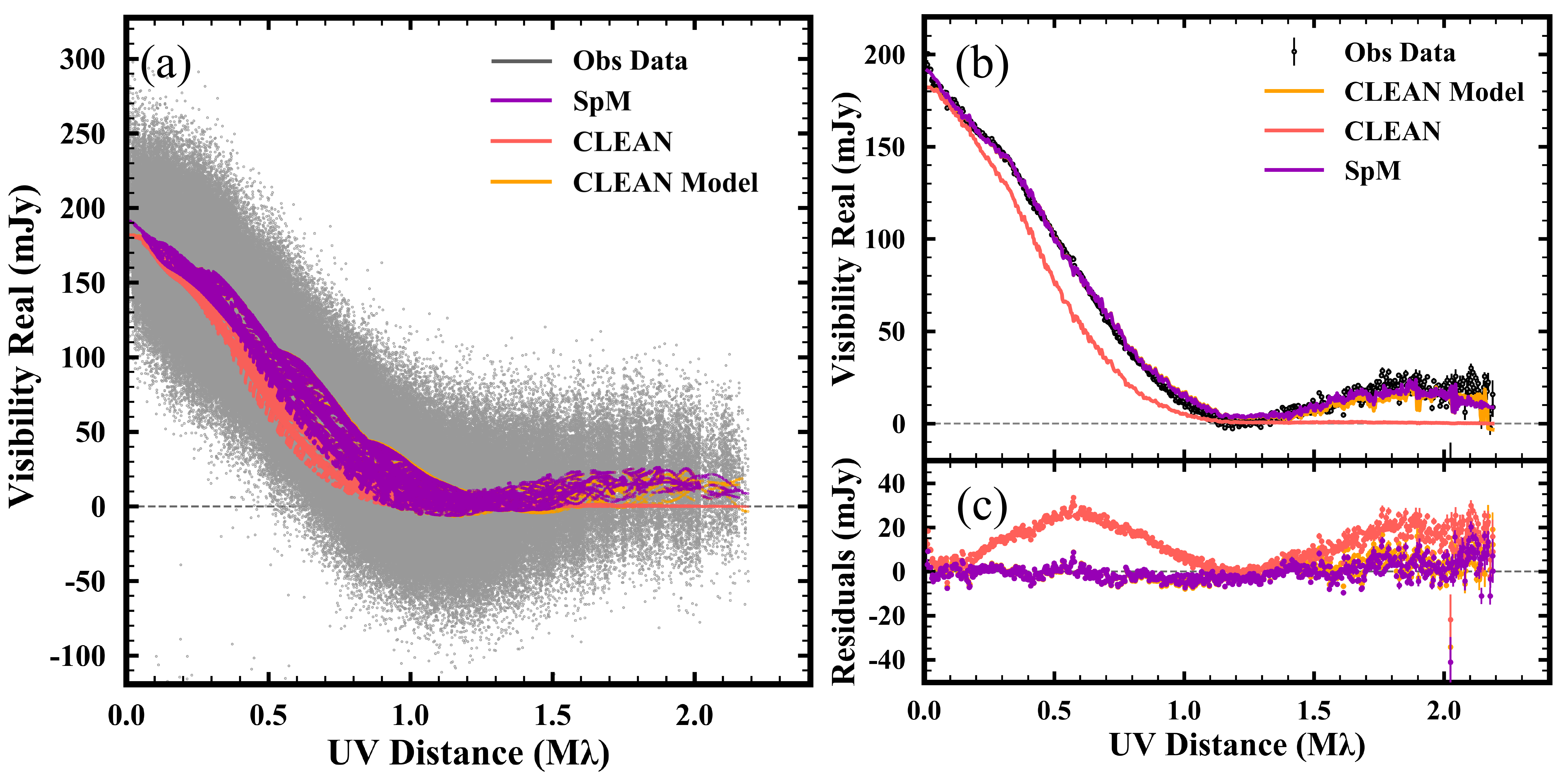}
\caption{(a) radial visibility profile gathered from radial visibilities in each azimuthal angle after deprojection in the $uv-$plane. This panel shows real parts of the visibilities for the SpM image (purple), the CLEAN image (rad), and the CLEAN model image (orange) together with the observational data (gray). (b) Binned and deprojected visibility profile in $3~\rm k \lambda$ bins. (c) Residual visibilities between each model and the observations.}
\label{fig:vis_profile}
\end{figure*}

Figure \ref{fig:vis_profile} shows radial visibility profiles of the T Tau system calculated from the SpM image, the CLEAN model, and the beam-convolved CLEAN image together with the real part of observed visibilities. Here, the visibility of the beam-convolved CLEAN image is obtained from the Fourier transform of the final CLEAN image (see Fig.\ref{fig:cont_images}(c)) by extracting the Fourier component that corresponds to the observed $uv$-sampling. The observed visibilities $O$ and modeled $M$ are deprojected in the $uv$-plane using the derived PA and inclination of the T Tau N disk in Appendix \ref{appendix:disk_geometry}.

To evaluate the goodness of fits between the models and the observation, we calculated the reduced-$\chi^{2}$. The formula is given by $\chi_{red}^{2}=N^{-1} \sum_{i=1}^{N}fW_{i}\left|O_{i}-M_{i}\right|^{2}$, where $N$ is the total number of the visibilities and $W_{i}$ is the weight of the $i-$th observed visibility $O_{i}$. The values of $W_{i}$ are obtained in the measurement set of ALMA data. The factor $f$ is the ratio between the weight and the standard deviation (stddev) of the visibility ($f=$stddev$^{-2}$/weight), which is reported to be $\sim 0.2-0.3$ in other disk observations \citep{Hashimoto2021ZZTau, Hashimoto2021DMTau}. To estimate $f$ of the T~Tau data, we calculated the standard deviation of the real part of visibility in every $3~k\lambda$ bins along the $uv$ distance. We have obtained that $f=0.29$. Finally, all the visibility models corresponding to the three images (CLEAN image, CLEAN model, and SpM) have resulted in the reduced $\chi^2$ of around unity: 1.31 for the CLEAN image, 1.18 for the CLEAN model, and 1.18 for the SpM, respectively.  Therefore, all the three images equally reproduces the visibility distribution in 2D $uv$-plane.

However, the situation changes when we consider azimuthally averaged visibility profiles. We have binned the visibility data every $3~k\lambda$ of the $uv$-distance and have taken average in each bin (Figure \ref{fig:vis_profile} (b, c)). The noise of the azimuthally averaged visibility is much smaller than the original 2D visibility. As a result, we found following two features; (1) The CLEAN model and the SpM image reproduce the observed visibility even after azimuthal average. (2) The visibility profile obtained from the CLEAN image significantly deviates from the observed visibility at $\rm 0.2-1.1~M\lambda$ and at $\rm 1.5-2.2~M\lambda$. We expect that the deviation of the CLEAN image is caused by the convolution by the restoring beam. We therefore consider that either the CLEAN model or the SpM image better reproduces observations compared to the CLEAN image.

It is not possible to distinguish the CLEAN model and SpM image from the goodness of fit measured by the reduced-$\chi^2$ values. However, we consider that the SpM image better reconstructs the disk surface brightness distribution. The CLEAN model reconstructs an image with a sum of a number of point sources \citep[CLEAN components;][]{hogbom1974, clark1980}. As a result, we see a patchy pattern in the CLEAN model image, which we consider irrelevant for disk structures. The SpM image shows more smooth structures than the CLEAN model and therefore seems more reasonable. Therefore, in this paper, we mainly use the SpM image for image analyses.  We do not yet have more quantitative measurements that can distinguish the SpM image from the CLEAN model image , and the bias that the SpM image may have is still an open question.

%-----------------------------Start Figure------------------------------
%-----------------------------End Figure---------------------------------

\section{Selection of Optimum Image in SpM Imaging of T Tau N Disk}\label{appendix:cvimage}

\begin{figure*}[t]
\centering
\includegraphics[width=0.99 \textwidth]{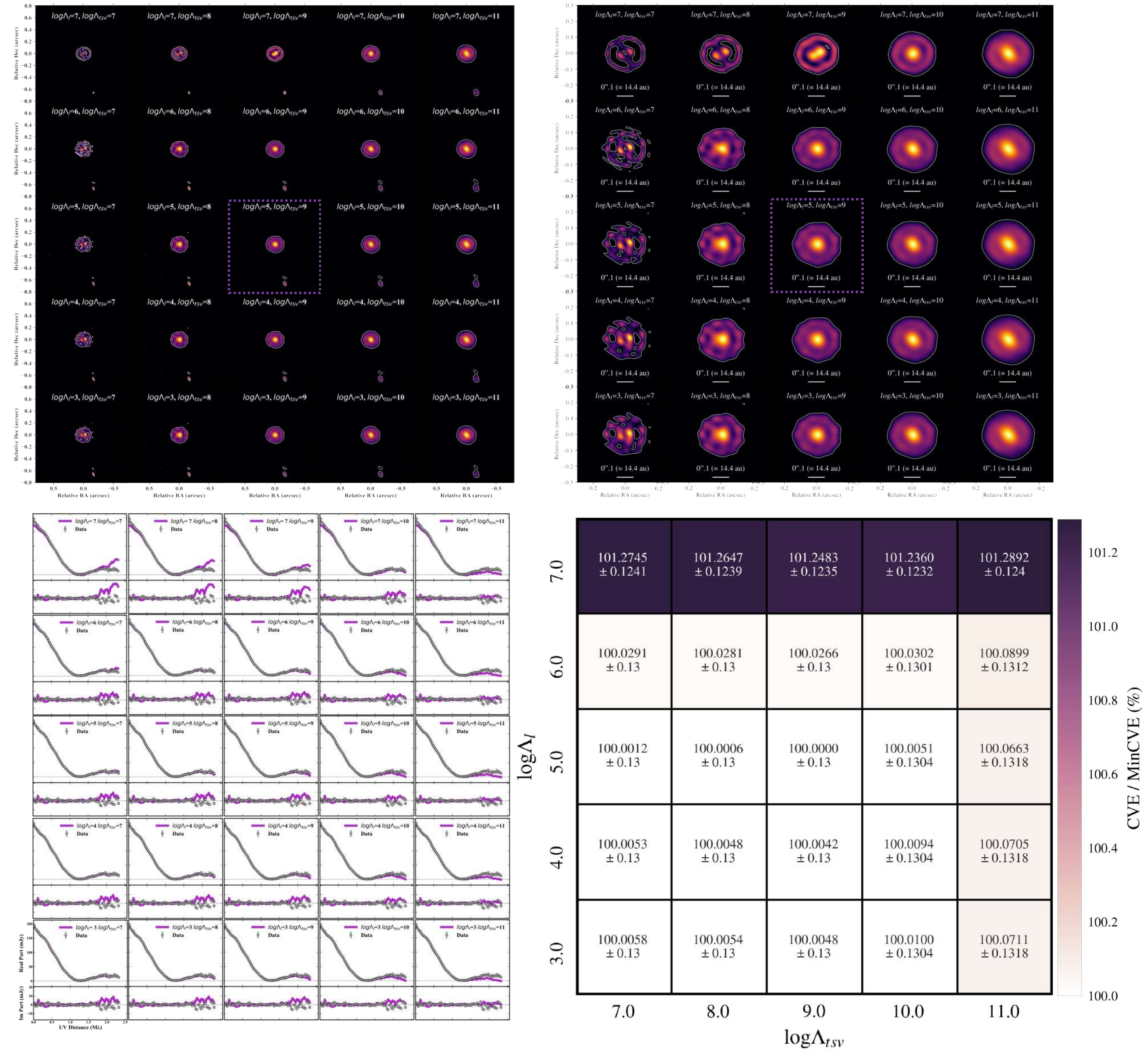}
\caption{SpM imaging of PPD T Tau system. Each panel corresponds to a gallery of 20 images with a combination of $\Lambda_{l}$ and $\Lambda_{tsv}$. (upper left): SpM images. A wide field of view of $1''.6\times1''.6$ is adopted. The dashed line box indicates the optimal image $\left(\Lambda_{l}=10^{5},~\Lambda_{tsv}=10^{9}\right)$ selected by CV. (upper right): Close-up view of SpM images. A field of view of $0.''5\times0.''5$ is adopted. (lower left): Radial-visibility profile The upper and lower panels indicate the real and imaginary parts obtained from the observed data (gray color) and the Fourier transform of the SpM images (purple color). (lower right): CV errors (CVEs) and $1\sigma$ uncertainties. The CVEs are the residual values between the observed data and SpM data using the mean squared error. Outputs denote the CVEs normalized to the minimum value.}
\label{fig:cvimage}
\end{figure*}

Figure \ref{fig:cvimage} shows 25 SpM images ($1''.6\times1''.6$) of the T Tau system and a closeup of T Tau N ($0''.5\times0''.5$), each of which corresponds to two sets of 25 SpM images of T Tau system corresponding to 25 combinations of regularization parameters $\left(\Lambda_{l},~\Lambda_{tsv}\right)$. We observe that the reconstructed disk structures change depending on the combination. We also show the visibility (real and imaginary parts) plots and calculated cross-validation error (CVE) with $1\sigma$ uncertainty. The optimal image is selected as one for $\left(\Lambda_{l}=10^{5},~\Lambda_{tsv}=10^{9}\right)$, giving the minimal CVE ($100.0000 \pm 0.13\%$). It is clearly seen that images with larger CVEs, i.e., $\left(\Lambda_{l}=10^{7}, \Lambda_{tsv}=10^{7}...10^{11}\right)$ and $\left(\Lambda_{l}=10^{3}...10^{7}, \Lambda_{tsv}=10^{11}\right)$ also show large deviations of model visibilities compared with observed data, especially at higher spatial frequencies. It should be noted that the annular gap structure of the T Tau N disk and the two separated emissions around T Tau Sa/Sb are commonly seen in images with a CVE of approximately $100.0\%$, for example, $\left(\Lambda_{l}=10^{4},~\Lambda_{tsv}=10^{9}\right)$ and $\left(\Lambda_{l}=10^{5},~\Lambda_{tsv}=10^{8}\right)$. This indicates that the presence of these structures is quite robust. 

In contrast, the image for $\left(\Lambda_{l}=10^{5},~\Lambda_{tsv}=10^{7}\right)$ also yields a low CVE ($100.0012 \pm 0.13\%$), and the inner disk is likely to be more resolved than other images. As described in Section \ref{sec:ImagingwithSpM}, the effective spatial resolution $\theta_{\rm eff}$ is $0''.03$. The ratio of the T Tau Sa disk size to the CLEAN beam was obtained as $\sim19\%$ for the image with $\left(\Lambda_{l}=10^{5},~\Lambda_{tsv}=10^{7}\right)$, which is smaller than $\theta_{\rm eff}$ ($\sim30\%$ of the CLEAN beam) for the optimal image with $\left(\Lambda_{l}=10^{5},~\Lambda_{tsv}=10^{9}\right)$. In a previous study, the SpM image allowed us to achieve a smaller beam size, that is, typically $\sim 30 - 40\%$ \citep{akiyama2017a,akiyama2017b, Kuramochi2018, yamaguchi2020} compared to the corresponding CLEAN image. The SpM image with such a ``hyper'' spatial resolution (i.e., beam smaller than $20\%$ of the CLEAN beam) may reflect the presence of a small substructure in the inner disk, but it appears difficult to evaluate the feasibility. Therefore, we conclude that the optimal image selected from the cross-validation (CV) would be the best among the images in Figure \ref{fig:cvimage} in terms of spatial resolution improvement, and it is also the best for quantitative analysis.

%-----------------------------Start Figure------------------------------
%-----------------------------End Figure---------------------------------

\section{Effective Spatial Resolution and Detection Threshold of SpM image}\label{appendix: effect_resolution}

\begin{figure*}[t]
\centering
\includegraphics[width=0.98 \textwidth]{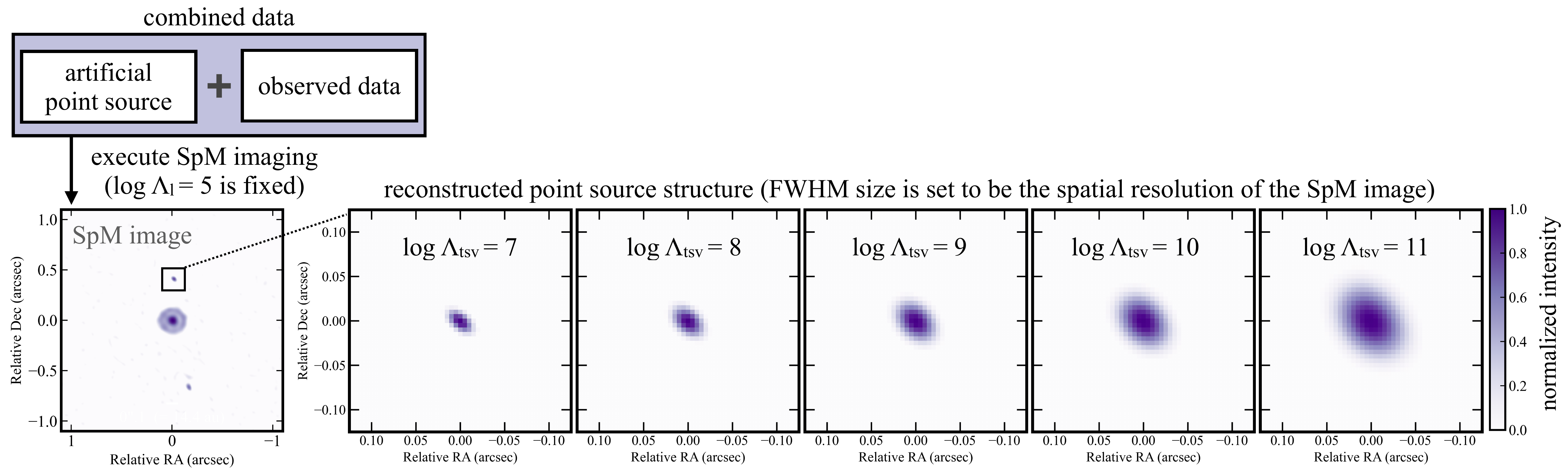}
\caption{Evaluation of effective spatial resolutions of the SpM images. The effective spatial resolution of the SpM image is evaluated in the way of an elliptical Gaussian fit to an artificial point source injected into $0''.4$ north in the observed data. Each resolution is tuned by the regularization parameter of $\Lambda_{tsv}$ while the regularization parameter of $\Lambda_l$ is fixed to be log $\Lambda_l = 5$.}
\label{fig:spm_resolution}
\end{figure*}

\begin{figure*}[t]
\centering
\includegraphics[width=0.98 \textwidth]{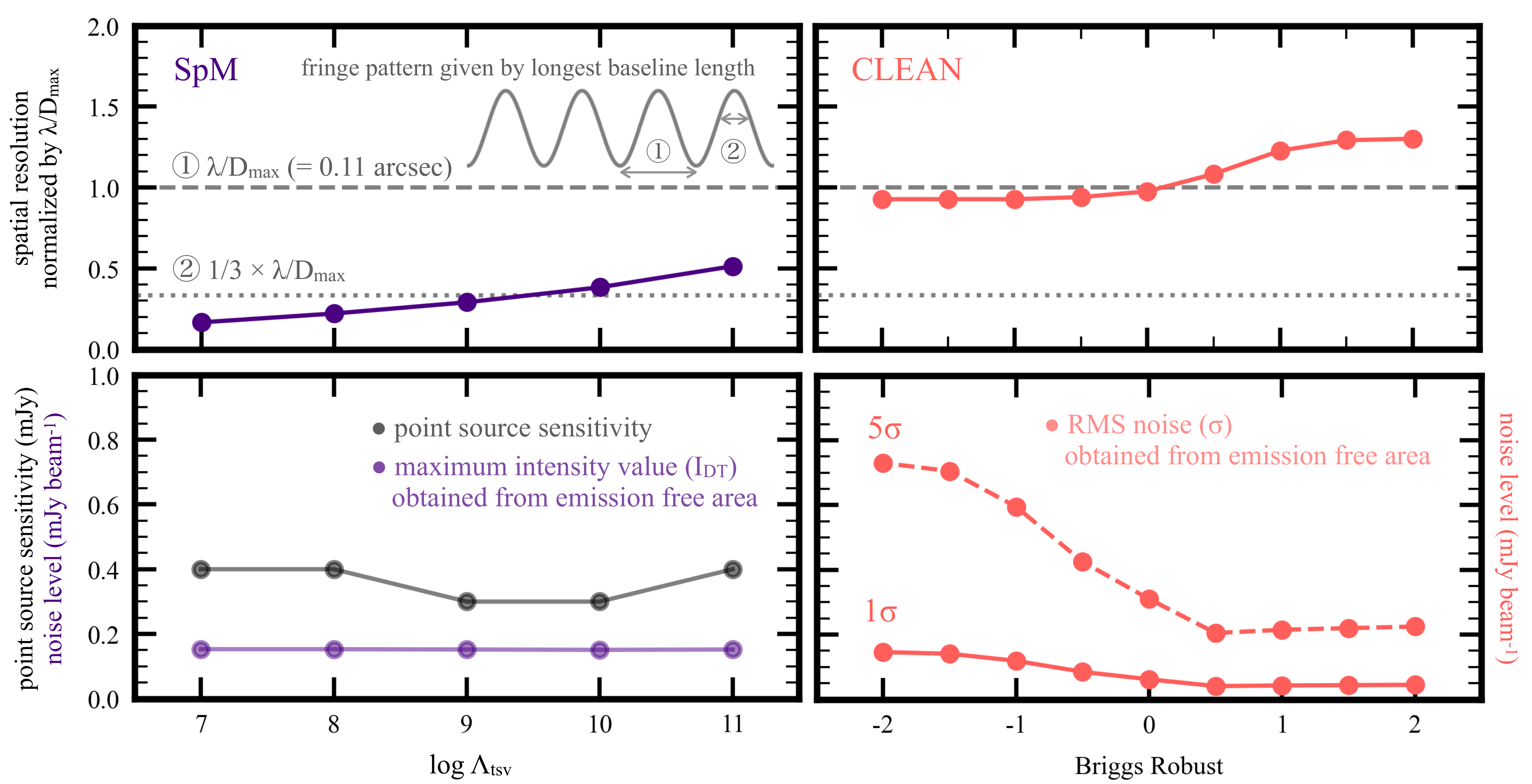}
\caption{Comparison of spatial resolutions and noise levels on the SpM (left side) and CLEAN (right side) images. Each resolution is plotted in the top panel as the geometric mean of the major and minor axes of beam (or point source, see Fig \ref{fig:spm_resolution}) normalized by the diffraction-limited resolution ($\lambda/D_{\rm max} = 0''.11$) given by the maximum baseline length $D_{\rm max}$. The left bottom panel indicates the point source sensitivity (gray color) in the SpM image set to be the threshold that the point source can be seen at the injected position with having a single peak with roughly more than $90~\%$ of the input flux density. The results are summarized in the panel together with $I_{\rm DT}$ (purple color) estimated from beam-convolved image (Jy/beam, beam size; $\theta = 0''.14 \times 0''.10$). The right bottom panel shows the RMS noise $\sigma$ of the CLEAN image for each $\tt Brigss~robust$ parameter.}
\label{fig:spm_clean}
\end{figure*}

We performed two kinds of SpM imaging simulations. One is for estimating effective spatial resolution in the SpM imaging, and another is for evaluating the detection threshold. For both, we injected an artificial point source at $0''.4$ north in the observed data. We made the SpM images for $\Lambda_{tsv} = 10^{7}, 10^{8}, 10^{9}, 10^{10}$, and $10^{11}$ and fixed $\Lambda_{l} = 10^{5}$. For the effective spatial resolution purpose, we injected the point source with a flux density of $7.1$ mJy (which corresponds to the total flux of the T Tau Sa disk). We fitted a two-dimensional Gaussian to the retrieved image of the point source. The obtained geometric mean of the source size of major and minor axis for each regularization parameter is plotted in Figure \ref{fig:spm_resolution}. For detection threshold estimate, we change the flux density of the point sources, from  $100~\mu \rm Jy$ to $1000~\mu \rm Jy$ in an increment of $100~\mu \rm Jy$. We judged the detection in the image according to the criterion that the point source can be seen at the injected position with having a single source with roughly more than $90~\%$ of the input flux density. The results are summarized in Figure \ref{fig:spm_clean} together with $I_{\rm DT}$ estimated from beam-convolved image (Jy/beam, beam size; $\theta = 0''.14 \times 0''.10$). The RMS noise $\sigma$ and CLEAN beam size for each robust parameter are summarized for comparison (see Figure \ref{fig:spm_clean}). The values of detection threshold for simulation in different $\Lambda_{tsv}$ or $I_{\rm DT}$ can be compared with CLEAN cases, and we found that those correspond to roughly $4\sigma$ of the CLEAN image for $\tt robust = 0.5$ and are lower than $4\sigma$ level for $\tt robust = -2$ or $-1$. These results show that the SpM imaging can achieve super-resolution without significant degradation of point-source sensitivity.

We then try to explain why the SpM imaging can achieve roughly three times better spatial resolution. The usual diffraction-limited resolution is roughly wavelength $\lambda$ divided by maximum baselines ($\lambda/D_{\rm max}$, where $\lambda$ is observing wavelength and $D_{\rm max}$ is an aperture size or the longest baseline length in interferometer). In the interferometric synthesis observations, the synthesized beam (i.e., response to a point-source or point spread function) can be expressed as a summation of each visibility pattern given by a two-element interferometer. The (one-dimensional) visibility pattern produced by the longest baseline has a fringe spacing of $\sim \lambda/D_{\rm max}$ \citep{thompson2017}. It then has a sharper spatial amplitude response with an FWHM of $ 1/3 \times \lambda/D_{\rm max}$ if we consider only its positive side contributing to forming the final beam. It is roughly three times smaller than the synthesized beam with a size of $\lambda/D_{\rm max}$ (see inset in Figure \ref{fig:spm_clean}). We have confirmed the FWHM size of $\sim 0''.04$ for the visibility pattern for the longest baseline ($\sim4$ km) in the ALMA configurations, DV23$\&$DA42. The SpM imaging is a regularized least-squares method where observed visibilities are directly used to retrieve the image, and the long baselines can be exploited as much as possible in super-resolution imaging. On the other hand, visibility amplitude distribution as a function of $uv$-distance can be used for estimating the source sizes of compact objects. For a source with Gaussian spatial distribution, the FWHM source size  $\Theta_{\mathrm{FWHM}}$ and the $uv$-distance giving the half visibility amplitude $UV_{1/2}$ have a relation such as, $\left(U V_{1 / 2}/100~k \lambda\right) \cdot\left(\Theta_{\mathrm{FWHM}}/1~\mathrm{arcsec}\right)=0.91 $ \citep[][]{Kawabe2018}. If the $UV_{1/2}$ is equal to $D_{\rm max}$, $\Theta_{\mathrm{FWHM}}$ in unit of radian is equal to $0.44 \times \lambda/D_{\rm max}$. It should be roughly $1/3 \times \theta_{\rm synth}$ if $\theta_{\rm synth}\simeq 1.22 \lambda/D_{\rm max}$ (corresponding to the Rayleigh criterion to resolve two-point sources; see ALMA technical handbook). This expression means that interferometric imaging could exploit visibility amplitudes measured with high SNR even at long baselines for deriving source structures much smaller than the synthesized beam.

%-----------------------------Start Figure------------------------------

%-----------------------------End Figure---------------------------------

%-----------------------------Start Figure------------------------------

%-----------------------------End Figure---------------------------------

\section{Disk Inclination and Position Angle of T Tau N}\label{appendix:disk_geometry}

\begin{figure}[t]
\centering
\includegraphics[width=0.48 \textwidth]{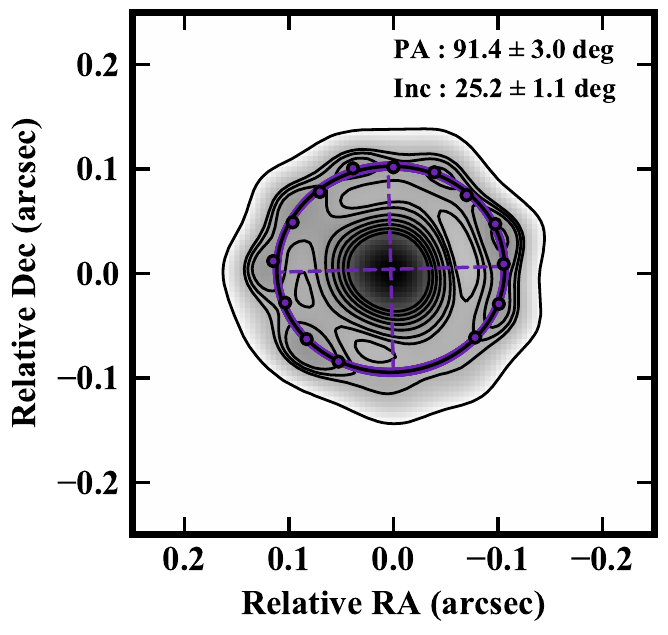}
\caption{Result of the best-fit ellipse to the outer ring overlaid on the SpM image. The black ellipse shows the best-fit model. The purple curves show the distribution of solutions considering the estimated $1\sigma$ error, while the purple dots along the outer ring indicate the radial averaged peak positions with an azimuthal angle spacing of $20^{\circ}$. The outer peak in the range of PA $=165^{\circ}\sim220^{\circ}$ cannot be identified because of an insufficient angular resolution or low signal-to-noise ratio, and it was excluded from the fitting. To clearly identify the silhouette of the outer ring, the image contour levels of $[1,8,9,10,12,14,16,18,20] \times I_{\rm DT}$ were adopted.}
\label{fig:ellipse_fit}
\end{figure}

To derive the position angle (PA) and inclination of the T Tau N disk using the SpM image, we performed an ellipse fit to the outer ring of the disk using least-squares fitting \citep{hammel2020} together with the Monte Carlo routine, as shown in Figure \ref{fig:ellipse_fit}. Here, we assume that the outer ring is a perfect circle in face-on, and we estimated the inclination angle from the aspect ratio of the ellipse. 

First, we sampled the radial peak position of the outer ring on the PA profile every $1^{\circ}$ in the azimuthal angle $\theta$. We averaged the peak position $r_{\rm peak}(\theta)$ and derived its standard deviation $\sigma_{\rm r}(\theta)$ from the measurements within the azimuthal angle spacing $\Delta\theta$, and used them for the ellipse fit. $\Delta\theta$ is set to be of the order of the major axis of the spatial resolution element $\theta_{\rm eff, maj}$ ($= 0''.038$) in Section \ref{sec:spmimage}. Given that the radial peak positions are roughly located at a radius of $r=0''.1$ (estimated from visual inspection), the azimuthal angle spacing in degrees is thus derived as $\Delta\theta \sim  \theta_{\rm eff, maj} \times 360^{\circ}/ 2\pi r \simeq 20^{\circ}$.

Because the radial peak positions in the range of PA = $165^{\circ} - 220^{\circ}$ cannot be identified owing to an insufficient angular resolution or low signal-to-noise ratio, these samples are excluded from the use of the ellipse fit. To calculate the error of the fit, a Monte Carlo routine was performed by randomly sampling the $\sigma_{\rm r}(\theta)$ deviation, and then $r_{\rm peak}(\theta)$ were added to them. 5000 iterative calculations were performed, and the best-fit ellipse can be obtained from the average values of the iterations. Uncertainties in each parameter of the ellipse fit were calculated by taking the standard deviation found with the iterations. The best-fit results of the PA and inclination are $91.4\pm3.0^{\circ}$ and $25.2\pm 1.1^{\circ}$, respectively, as summarized in Table \ref{tab:TTauN_disk}.

%-----------------------------Start Figure------------------------------
%-----------------------------End Figure---------------------------------

\section{Spectral Index of T Tau N Disk}\label{appendix:spectral_index_TTauN}
The spectral index $\alpha$ of the T Tau N disk was estimated using two spectral windows (spws) that were used for the continuum observations at Band 6, with center frequencies at  $\nu_{1}$ = 218 GHz and $\nu_{2}$ = 233 GHz. Two CLEAN images were restored with the same beam size identical to that obtained at a lower frequency. Using the CLEAN maps, we obtained the total flux densities, $F_{1}$ and $F_{2}$ for $\nu_{1}$ = 218 GHz and $\nu_{2}$ = 233 GHz, respectively. The spectral index was calculated as $\alpha$ = ln($F_{1}/F_{2}$) / ln($\nu_{1}/\nu_{2}$), and was obtained as $\alpha = 1.9 \pm 0.1$. For the optically thick emission, the Rayleigh-Jeans limit gives $\alpha = 2$. For other PPDs in the Taurus, the typical $\alpha_{\rm mm}$ values are below 3.0 and several of them are even below 2.0, as in the case of the T Tau N disk \citep[e.g.,][]{Ricci2010, Akeson2014, Pinilla2014, Ribas2017, Zagaria2021}. The obtained spectral index lower than the limit can be explained by considering the additional effect of dust self-scattering \citep{baobab2019, zhu2019}. 

% Bibliography
\bibliographystyle{yahapj}
\bibliography{ref}
\end{document}